\documentclass[twocols]{aa}  
\usepackage{linenoaa}
\usepackage{float}
\usepackage{makecell}
\usepackage{xcolor}
\usepackage{graphicx, booktabs}
\usepackage{txfonts}
\usepackage{placeins}

\begin{document} 
\nolinenumbers

   \title{Energetics of star--planet magnetic interactions}

   \subtitle{Novel insights from 3D modelling}

\author{Arghyadeep Paul\inst{1}, Antoine Strugarek\inst{1}
          }

   \institute{Université Paris Cité, Université Paris-Saclay, CEA, CNRS, AIM, F-91191, Gif-sur-Yvette, France\\
              \email{arghyadeepp@gmail.com / arghyadeep.paul@cea.fr}
             }
\abstract{Star--planet magnetic interactions (SPMIs) occurring in the sub-Alfvénic regime can, in principle, induce stellar chromospheric hotspots. These hotspots could serve as observational markers for inferring key planetary properties, especially the exoplanetary magnetic field, which is otherwise notoriously difficult to constrain. Currently, estimates of the power generated by SPMIs primarily rely on analytical scaling laws that relate stellar and planetary parameters to the interaction energetics. The existing scaling laws published in the literature so far do not agree with each other by at least an order of magnitude.}
{Our aim was to quantify an absolute upper limit on the power that a planet can channel back to its host star during such interactions, which in turn can lead to the formation of stellar hotspots. Furthermore, we explored how this energy varies with different planetary characteristics and the stellar wind conditions in the vicinity of the planet.}
{We employed three-dimensional numerical simulations of SPMIs in which the planet orbits within the sub-Alfvénic regime of the stellar wind. By performing a series of simulations with varied parameters known to influence the energetics of SPMIs, we derive a numerically supported scaling law that can be used to reliably estimate the energy channeled from the planet back to the star.}
{Our results suggest that existing analytical scaling laws may not fully capture the power transferred from the planet to the star through SPMI. The scaling law derived from our numerical simulations appears to provide a more comprehensive estimate, reflecting dependences on common stellar and planetary parameters also considered in earlier models. Moreover, our findings indicate that power generation involves not only the planetary obstacle itself but also the extended magnetic structure of the Alfvén wings interacting with the streaming stellar wind.}
{This study suggests that care should be taken when applying analogies directly from Jovian sub-Alfvénic interactions to SPMIs, as the underlying physical conditions (specifically the value of the Alfvénic Mach number) may not be directly comparable. Our numerically derived scaling law offers a potentially improved approach for estimating SPMI power, capturing some of the interaction’s complexities exclusive to SPMIs.} 


   \maketitle
%
\nolinenumbers
\section{Introduction}
Exoplanets undergo a wide range of interactions with their host stars, that can significantly influence their atmospheres, magnetic environments, and orbital dynamics \citep{Vidotto_2019, Strugarek_2024, Gourves_2025}. One of the most prominent forms of these interactions is known as star--planet magnetic interaction (SPMI), particularly relevant for exoplanets in close-in orbits. During SPMIs, magnetic coupling creates flux tubes that tether the planet to the star, allowing energy exchange in both directions \citep{Strugarek_2018, Fischer_2022}. This magnetic connection is only possible if the planet lies within the star’s Alfvén surface, a three-dimensional boundary around the star where the stellar wind speed matches the local Alfvén speed \citep{Strugarek_2022, Vidotto_2023}. Planets located outside this surface remain magnetically disconnected from the star as the wind speeds exceed the Alfvén speed. In our Solar System, all planets orbit beyond the Sun’s Alfvén surface; however, sub-Alfvénic interactions do occur within planetary magnetospheres, such as the interactions between Jupiter and its moons Io, Europa, and Ganymede \citep{Saur_2013}. Much of our current understanding of sub-Alfvénic star–planet interactions has been inferred from observations of Jovian analogs.
 \begin{figure*}
    \centering
    \includegraphics[width=1\linewidth]{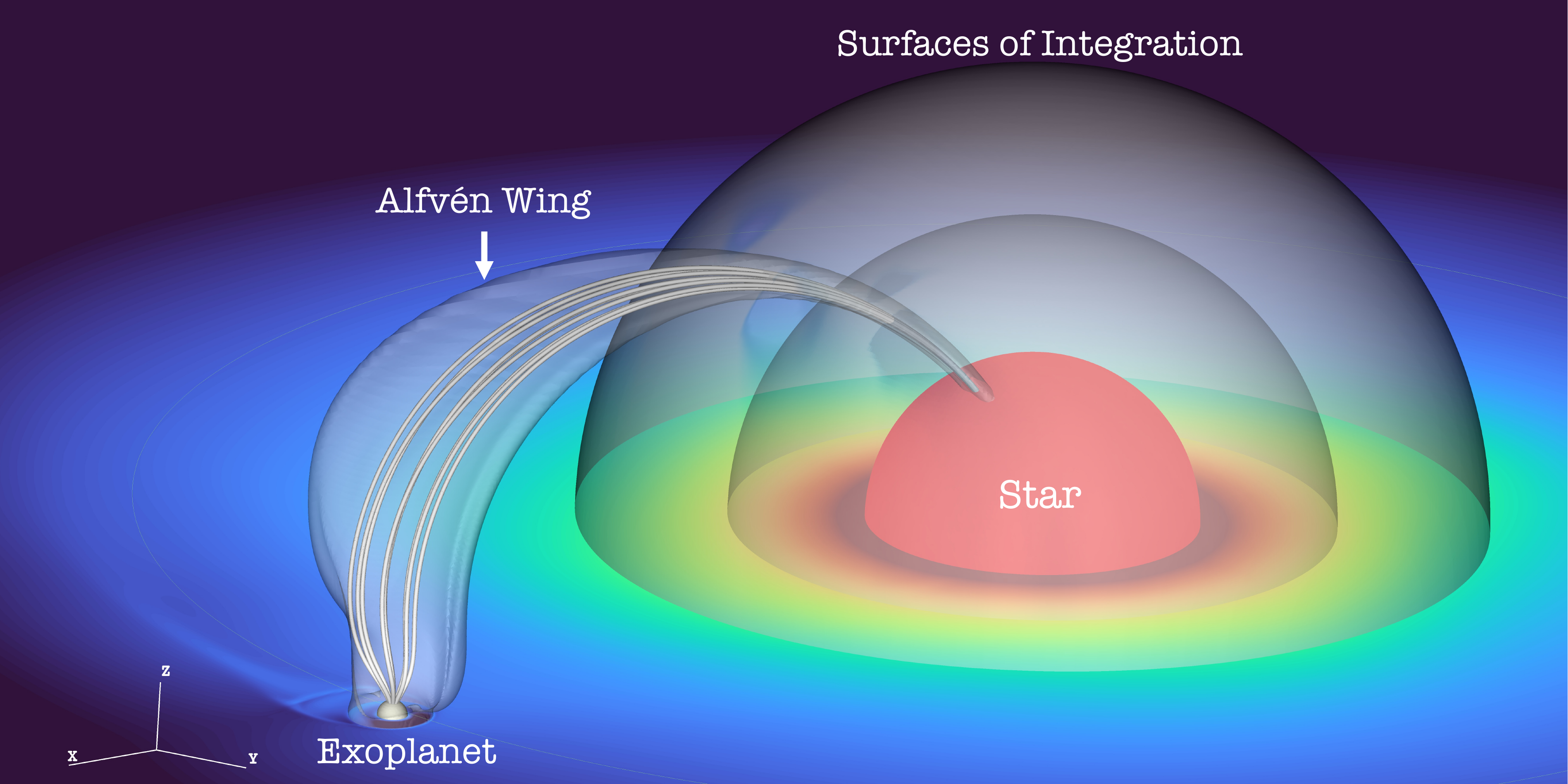}
    \caption{Schematic from a simulation depicting a close-in exoplanet orbiting its host star. The central red sphere represents the star, while the smaller white sphere indicates the exoplanet. White tubes illustrate magnetic field lines connecting the planet and star, with the surrounding sheath representing the isosurface of one of the Alfvén wings. Translucent hemispheres show the surfaces over which inward Poynting flux was integrated (see main text for details).}
    \label{fig:3d_schematic}
\end{figure*}
After the seminal prediction of the existence of SPMIs \citep{Cuntz_2000,Ip_2004}, the first tentative evidence for sub-Alfvénic SPMIs came from observations by \citet{Shkolnik_2005} and \citet{Shkolnik_2008}, who found that the stars HD 179949 and $\upsilon$ Andromedae, each hosting a hot Jupiter, displayed enhanced chromospheric activity synchronized with the planet’s orbital period. Follow-up studies such as \citet{Cauley_2019} supported this connection in a few systems and estimated the energy output from these chromospheric hotspots. However, it has been shown that such activity indicators can vary significantly over time, with signals appearing during some epochs and disappearing during others. This variability is evident in observations of HD 189733 \citep{Cauley_2018}, as well as in numerical simulations presented by \citet{Strugarek_2022}.

Current theories suggest that close-in exoplanets within the Alfvén surface can produce substantial energy flux through several mechanisms. These include magnetic reconnection between stellar and planetary fields \citep{Cuntz_2000}, the generation of Alfvén waves as the planet obstructs the stellar wind \citep{Saur_2013}, and the dissipation of magnetic stresses induced by the planet at the stellar chromosphere \citep{Lanza_2013}. As the planet moves through the sub-Alfvénic plasma, it carves out current structures known as Alfvén wings. These structures act as the aforementioned magnetic tethers connecting the planet and the star \citep{Fischer_2022, Strugarek_2015, Strugarek_2019}. Figure \ref{fig:3d_schematic} shows a portion of these wings, visualized as a translucent white isosurface of $\textbf{\textit{s}} \cdot \textbf{\textit{c}}_A$ (this nomenclature is defined below), from one of the simulations presented in this study.

The current understanding is that energy extracted from the local stellar wind near the planet is carried along magnetic field lines toward the star in the form of Alfvén waves. These waves propagate through the stellar corona and transition region, eventually dissipating in the chromosphere, where they produce hotspots that are synchronized to the planet’s orbital motion. \citep{Lanza_2012, Lanza_2013, Saur_2013, Cauley_2018, Cauley_2019, Strugarek_2015, Strugarek_2016, Strugarek_2018, Strugarek_2019, Shkolnik_2003, CastroGonzalez_2024}. 
These stellar hotspots are expected to form at the footpoints of Alfvén wings. Figure \ref{fig:3d_schematic} illustrates the configuration, where the Alfvén wing field lines are shown as white tubes connecting the planet to the star. It is at the stellar footpoints of these field lines that energy deposition is expected to occur, giving rise to localized regions of enhanced emission, commonly referred to as "stellar hotspots." However, several uncertainties remain in understanding this mechanism completely. For hotspots to form via this pathway, energy must first be generated in the vicinity of the exoplanet, then transferred along the Alfvén wings, and ultimately deposited in the stellar atmosphere. This process is governed by several efficiency factors \citep{Strugarek_schkolnik_2025}, which we outline below using tentative nomenclature. Four prominent efficiency factors exist: E1, which corresponds to the energy redirected along Alfvén wings from the locally available energy; E2, which accounts for propagation losses; E3, which corresponds to the energy reflected back from the stellar transition region; and E4, which determines the emission efficiency of the chromosphere. Recently, \citet{Paul_2025} provided constraints on E3, showing that in realistic SPMI systems only $\sim$10\% of the incident energy passes through the stellar transition region.

Analytical models have been proposed to derive scaling laws that estimate the power channeled by Alfvén wings back to the host star in the context of SPMIs. Among the most widely referenced models in the literature are those developed by \citet{Saur_2013} and \citet{Lanza_2013} (hereafter referred to as the Saur model and the Lanza model, respectively). These models are discussed in greater detail in Section \ref{sec:analytical_models}. The Saur model, in particular, has demonstrated notable success in reproducing the energy budget of sub-Alfvénic interaction observed between Jupiter and its magnetically connected moons, such as Io, Europa, and Ganymede. However, when applied to SPMI scenarios, the model tends to underpredict the power output by approximately an order of magnitude compared to observational inferences \citep{Saur_2013,Cauley_2018,Cauley_2019}. This discrepancy raises the important question of why a model that effectively captures the energetics of sub-Alfvenic interactions for the Jovian moons falls short when applied to expectedly analogous star--planet systems. This apparent inconsistency highlights the need for further investigation into the underlying assumptions and parameter regimes of existing models. \citet{Strugarek_2016} found that the Poynting flux within the Alfvén wings near the exoplanet for close-in interactions aligns with predictions derived from the Saur model. However, such an approach of considering just the flux contained within the Alfvén wings near the planet may not fully account for the total power ultimately transmitted to the star due to SPMIs. The Lanza model, while capable of predicting power outputs that are broadly consistent with the energy requirements of SPMIs from an order-of-magnitude perspective, has not yet been rigorously validated through numerical simulations. As a result, its applicability to more realistic, three-dimensional sub-Alfvénic star–planet interaction scenarios remains uncertain. It is therefore important to assess whether the power estimates derived from the Lanza analytical model remain consistent with those obtained from self-consistent 3D numerical simulations, which can capture the complex geometry and dynamics inherent to such systems. 

\citet{Cauley_2019} presented an important proof-of-concept study exploring the use of SPMI-induced hotspots to infer exoplanetary magnetic fields, with results broadly consistent with theoretical expectations from internal dynamo models \citep{Yadav_2017} for hot Jupiters. Building on this, \citet{Paul_2025} showed that including energy reflection at the stellar transition region can significantly increase the inferred field strengths, sometimes up to a few hundred gauss. These differences underscore the importance of considering additional physical effects in SPMI interpretations, adding complexity but also offering a path toward more comprehensive inferences.

For this study we focused on addressing one such fundamental and yet unresolved question in the field: how much power SPMIs can generate, for a planet orbiting within the sub-Alfvénic regime of a stellar wind. Our goal is to determine the scaling behavior of the power produced by SPMI and to evaluate how it compares with the currently established analytical models in the literature.To achieve this, we employed three-dimensional numerical simulations of a set of representative SPMI systems based on the initial study of \citet{Strugarek_2016}, varying key parameters that are expected to influence the nature and strength of the interactions. Through these simulations, our aim was to gain insights into the underlying physics and to assess the analogy of the Jovian sub-Alfvénic interactions in relation to SPMI systems. 

The structure of the paper is as follows. Section \ref{sec:num_setup} describes the numerical setup used in this work, while Section \ref{sec:analytical_models} presents the analytical models commonly used as reference points and lays the foundation for comparison with our numerical results. Sections \ref{sec:alf_wing_str_en_flx} and \ref{sec:SPMIpowerestimates} present and discuss the key results obtained from this study. Finally, Section \ref{sec:disc_and_conc} concludes the study and offers important discussions that help interpret our findings in the context of current observational constraints on SPMIs.

\section{Numerical setup\label{sec:num_setup}}
We use the open-source PLUTO code \citep{Mignone_2007} to numerically solve the ideal magnetohydrodynamic (MHD) equations and simulate star–planet magnetic interactions (SPMIs). The goal is to characterize the total energy flux that a close-in planet can transfer to its host star through these interactions. Our simulation setup is largely based on the model developed by \citet{Strugarek_2015}, which we briefly summarize below for clarity. PLUTO solves the following set of ideal MHD equations:

\begin{align}
  \partial_t \rho  + \nabla \cdot(\rho \textbf{v}) &= 0 \\
  \rho \partial_t \textbf{v} + \rho \textbf{v}\cdot \nabla \textbf{v} + \nabla P + \textbf{B} \times \nabla \times \textbf{B} / \mu_0  &= \rho \textbf{a}\\
  \partial_t P + \textbf{v}\cdot \nabla P + \rho c_s^2 \nabla \cdot \textbf{v} &= 0 \\
  \partial_t \textbf{B} - \nabla \times (\textbf{v}\times\textbf{B}) &= 0\\
  \nabla \cdot \textbf{B} &= 0
\end{align}
In these equations, $\rho$ represents the plasma mass density, $\textbf{v}$ is the bulk velocity, $P$ is the gas pressure, and $\textbf{B}$ denotes the magnetic field. The source term $\textbf{a}$ in the momentum equation includes contributions from gravitational, centrifugal, and Coriolis forces. The equations are solved in a rotating reference frame that co-rotates with the planet's orbital motion, effectively keeping the planet stationary within the computational domain. The adiabatic sound speed is given by $c_s = \sqrt{\gamma P / \rho}$, where $\gamma$ is the adiabatic index. The system is closed using the ideal gas equation of state, $\rho \varepsilon = P / (\gamma - 1)$, with $\varepsilon$ denoting the internal energy per unit mass.

The simulation setup comprises two main components: the stellar wind model and the planetary model. The stellar wind is simulated using the Wind-Predict framework from \citet{Reville_2016}, while the planet is modeled as a magnetized body with an electrostatic ionosphere, following the MagPIE framework \citep{Paul_2023}. Both models are implemented within the PLUTO code and coupled into a unified simulation environment. For simplicity, the planetary ionosphere is assumed to be a perfect electrical conductor. The magnetic fields of both the star and the planet are represented as dipolar. The computational domain extends from –40 to 40 $R_{\star}$ in each of the three spatial directions (X, Y, Z). The grid employs a combination of uniform and stretched regions to balance resolution with computational efficiency. In the X-direction, the grid consists of 375 cells in total, including a uniform segment of 97 cells spanning –1.5 to 1.5 $R_{\star}$ to resolve the star, with a resolution of $\Delta X = 0.03 R_{\star}$. An additional uniform segment of width $1 R_{\star}$ is placed at the planet’s location to resolve the interaction region with 81 cells, corresponding to a finer resolution of $\Delta X = 0.01 R_{\star}$. The Y and Z directions each contain 275 cells in total and feature a central uniform region from –1.5 to 1.5 $R_{\star}$, matching the resolution used in the X-direction for the star. Around the Y and Z positions of the planet, the grid further refines to the same fine resolution as in X, ensuring cubical grid cells at the refined region near the planet. Outside these uniform regions, the grid transitions smoothly into stretched zones, where the cell size increases gradually with distance from the uniform–stretched interface. The stretched regions begin with cell spacing equal to that of the adjacent uniform region and expand progressively toward the boundaries of the domain.

A series of simulations is performed, varying the planetary orbital radius and magnetic field strength, with the configurations detailed in Table~\ref{tab:simulation_setups}. The quantity $R_{orb}$ represents the orbital distance of the planet whereas $B_p$ represents the equatorial surface magnetic field strength of the planet. The magnetic topology at the planetary magnetospheric boundary is determined by the relative orientation of the stellar and planetary magnetic dipole moments. Depending on the sign of this alignment, the system can adopt aligned, anti-aligned, or intermediate magnetic configurations. These represent two limiting magnetospheric states: a “closed” magnetosphere when the dipole moments of the star and planet are aligned, and an “open” magnetosphere in the anti-aligned case. \cite{Ip_2004} illustrates these open and closed magnetospheric configurations resulting from the two opposing dipolar orientations, specifically in panels (a) and (c) of Figure 2 in \citet{Ip_2004}, which correspond respectively to the open and closed configurations discussed in this paper.

For completeness, we report the substellar magnetopause standoff distance ($R_{MP}$) from the center of the planet for the open-magnetospheric configurations for each of the simulated setups in table \ref{tab:simulation_setups}. The open configuration is expected to maximize the Poynting flux generated through SPMI. Accordingly, all baseline cases in Table~\ref{tab:simulation_setups} assume an open magnetosphere to capture this upper bound. For specific orbital radii, both open and closed configurations were simulated to have a conservative estimate on the lower bound; these are denoted as $(+/-)$ in the table. A change in the orbital radius of the planet influences the planet’s orbital speed, which in turn modifies its relative motion with respect to the stellar wind plasma. Simultaneously, the planet encounters different regions of the stellar magnetic field, leading to variations in the local stellar magnetic field strength at the vicinity of the planet. These two factors jointly influence the total power budget of SPMIs.

For all simulations in this study, corresponding background stellar wind runs, with only the stellar wind and excluding the planet, were also performed to establish baseline reference values for each analyzed quantity. These baseline values were subtracted from all the results of simulations that include the planet, thereby isolating the specific contributions attributable to the exoplanet. As a result, all quantities and plots presented in the following sections reflect background-subtracted values, highlighting only the exoplanetary contribution. In all simulated magnetospheric configurations (open and closed), the system evolves to form Alfvén wings. The Alfvén wings for an open magnetospheric case is illustrated in Figure~\ref{fig:3d_schematic}. This figure shows a representative setup, with the star depicted as the red sphere and the planet as the smaller white sphere. The magnetic field lines comprising or adjacent to the Alfvén wings act as magnetic tethers that connect the planet to the star, enabling the transfer of energy and information along the field. The Alfvén wing profile is shown as a curved white isocontour surface enclosing these magnetic field lines. In addition, the figure includes several spherical integration surfaces centered on the star, which will be discussed in more detail in the following sections, particularly in Section \ref{sec:SPMIpowerestimates}.

\begin{table*}[]
\centering
\begin{tabular}{lccccc}
\toprule\toprule
                     & $R_{orb}/R_{\star} = 3.0$ & $R_{orb}/R_{\star} = 3.9$ & $R_{orb}/R_{\star} = 5.0$ & $R_{orb}/R_{\star} = 5.9$ & $R_{orb}/R_{\star} = 7.0$ \\ 
\midrule
\midrule

$B_{p} = 2.0 \times 10^{-4}$ T & -- & \makecell{r3.9b1 \\ ($R_{MP}=~\text{0.21}~R_{\star}$)} & -- & -- & -- \\ 
\midrule
$B_{p} = 1.5 \times 10^{-4}$ T & \makecell{r3b2 \\ ($R_{MP}=~\text{0.16}~R_{\star}$)} & \makecell{r3.9b2 (+/-) \\ ($R_{MP}=~\text{0.19}~R_{\star}$)} & \makecell{r5b2 (+/-) \\ ($R_{MP}=~\text{0.26}~R_{\star}$)} & \makecell{r5.9b2 \\ ($R_{MP}=~\text{0.3}~R_{\star}$)} & \makecell{r7b2 (+/-) \\ ($R_{MP}=~\text{0.33}~R_{\star}$)} \\ 
\midrule
$B_{p} = 1.0 \times 10^{-4}$ T & -- & \makecell{r3.9b3 \\ ($R_{MP}=~\text{0.18}~R_{\star}$)} & -- & \makecell{r5.9b3 \\ ($R_{MP}=~\text{0.28}~R_{\star}$)} & -- \\ 
\midrule
$B_{p} = 5.0 \times 10^{-5}$ T & -- & \makecell{r3.9b4 \\ ($R_{MP}=~\text{0.15}~R_{\star}$)} & -- & \makecell{r5.9b4 \\ ($R_{MP}=~\text{0.22}~R_{\star}$)} & -- \\ 

\bottomrule
\end{tabular}
\caption{Grid of simulations performed. The (+/-) sign denotes that for this configuration, in addition to an open magnetosphere (-), a closed magnetosphere (+) has also been tested. The value of $R_{MP}$ shown in parentheses below each label indicates the substellar magnetopause standoff distance from the center of the planet for the corresponding open (-) configuration. The stellar equatorial dipolar magnetic field is fixed at 12.4 G in all simulations.}
\label{tab:simulation_setups}
\end{table*}

\section{Existing analytical approaches to SPMI\label{sec:analytical_models}}
To establish a reference for comparison, we evaluate our simulation results against two widely used analytical models for estimating the power generated by SPMI: the Saur model \citep{Saur_2013} and the Lanza model \citep{Lanza_2013}. Such models were originally introduced by \citet{Zarka_2001} and \citet{Zarka_2007} in the context of predicting exoplanetary radio emission (see \citet{Callingham_2024} for a review). The derivation of \citet{Saur_2013} captures a similar physical effect as \citet{Zarka_2007}, differing only by a factor of $2M_A$ \citep{Saur_2013}. For conciseness, we therefore compare our results only with the Saur and the Lanza model. To provide context for the comparisons presented in the following sections, we first offer a brief overview of these two models. Both models estimate the Poynting flux directed from the planet toward the star in the regime of sub-Alfvénic interactions.

To set the stage for the discussion that follows, we begin by introducing the general expression for the Poynting vector, which represents the flux of electromagnetic energy. It is defined in SI units as
\begin{equation}
\vec{s} = \frac{\vec{E}\times \vec{B}}{\mu_0}
\end{equation}
In the context of the Saur model, the effective obstacle size of a magnetized planet is estimated using the relation
\begin{equation}\label{eq:eff_obs_size}
\rm R_{obs} = R_p \left(\frac{B_p}{B_{w}}\right)^{\frac{1}{3}},
\end{equation}
where $R_p$ denotes the planetary radius, $B_p$ the planetary magnetic field, and $B_w$ the stellar magnetic field at the planet’s orbital location.
Based on this obstacle size, the corresponding power output can be expressed as the Poynting flux integrated over the cross-sectional area of the Alfvén wing:
\begin{equation}
\rm S_{Saur} = 2\pi R_{eff}^2 v_A \frac{(\alpha M_A B_{w} \cos \theta)^2}{\mu_0} \qquad [\text{watts}]. \label{eqn:Saur_orig}
\end{equation}
Here $v_A$ is the Alfvén speed; $M_A$ the Alfvénic Mach number defined as the ratio ($v_{rel}/v_A$), where $v_{rel}$ is the relative velocity between the obstacle and its surrounding plasma; $\alpha$ the interaction efficiency; $\theta$ is the angle between the local magnetic field and the direction of the ambient plasma flow; and $R_{\rm eff}$ is the effective obstacle size which has a maximum value of $\sqrt{3}R_{obs}$. In the original derivation by \cite{Saur_2013}, the expression of equation \ref{eqn:Saur_orig} is derived in the limit where $M_A \rightarrow 0$. For an upper bound estimate, we assume $\alpha = 1$ and $\cos \theta = 1$, yielding
\begin{equation}\label{eq:Saur_power}
\rm S_{Saur (Max)} = 6\pi R_{obs}^2 v_A \frac{(M_A B_{w})^2}{\mu_0} \qquad [\text{watts}].
\end{equation}

The Lanza model, often referred to in the literature as the ``stretch-and-break'' model, provides the total SPMI power as
\begin{equation}\label{eq:Lanza_power}
\rm S_{Lanza} = \frac{2 \pi f_{AP} R_p^2 B_p^2 v_{rel}}{\mu_0} \rm \qquad [watts],
\end{equation}
where $R_p$, $B_p$, and $v_{rel}$ denote the planetary radius, polar magnetic field, and relative velocity between the planet and stellar wind. Here, $v_{rel}$ is taken as the vector difference of the orbital and stellar wind speeds at the planet’s location. The quantity $f_{AP}$ represents the fraction of the planetary surface magnetically connected to the stellar field and is given by
\begin{equation}
    \rm f_{AP} = 1- \left( 1-\frac{3 \zeta^\frac{1}{3}}{2+\zeta}\right)^{\frac{1}{2}},
\end{equation}
where $\zeta$ is defined as
\begin{equation}
    \rm \zeta = \frac{B_{\star (r= r_{orb})}}{B_p}.
\end{equation}
Each of the equations \ref{eq:Saur_power} and \ref{eq:Lanza_power} represents the power along a single Alfvén wing; therefore, a factor of 2 is applied in subsequent calculations to account for the contributions from both the northern and southern Alfvén wings connecting to the star.

Recent tentative estimates of SPMI power, inferred from excess chromospheric emission, suggest values on the order of $10^{19}$ to $10^{20}$ watts. If such emission hotspots are indeed driven by energy deposited due to SPMI on the star, the interaction must supply at least $\sim 10^{20}$ watts to account for the observed enhancements. However, magnetic reconnection alone is expected to produce power levels nearly three orders of magnitude below these estimates. Explaining the observed SPMI power purely through reconnection would require unrealistically strong planetary magnetic fields, on the order of $10^4$ G.

Within the framework introduced by \citet{Saur_2013}, the Saur model posits that power is generated at the planet and transmitted to the star via Alfvén wings connecting the two bodies. This model has been successfully applied to the Jupiter–moon system, including Io, Europa, and Ganymede, yielding power outputs between $10^9$ and $10^{12}$ watts, values consistent with observed Jovian auroral hotspots for analogous sub-Alfvénic planet-moon interactions. However, when extrapolated to exoplanetary systems such as HD 179949b, the model requires a planetary magnetic field approximately three orders of magnitude stronger than Jupiter’s to reproduce observed SPMI power levels around $10^{20}$ watts \citep{Cauley_2019}. Furthermore, the Saur model generally underpredicts the required tentative SPMI powers by roughly an order of magnitude, even in optimistic configurations. These discrepancies suggest that directly applying Solar System analogs to exoplanetary SPMI systems may have limitations, and they highlight the importance of exercising caution when interpreting such models in the context of SPMIs.

The Lanza model, though not yet validated by numerical simulations, has remained the only existing analytical framework capable of producing SPMI powers on the order of $10^{20}$ to $10^{21}$ watts so far. This is particularly important given that the observed stellar hotspot emissions likely represent only a fraction of the total power generated by the SPMI. The model proposes that magnetic stresses induced by the planet’s orbital motion are transmitted to the star along the connecting flux tube. It is also noteworthy that both the Lanza and Saur models include scaling parameters ($\alpha$, $f_{AP}$) that characterize the `efficiency' of the interaction between the stellar wind and the planet. These factors remain poorly constrained and can significantly affect the predicted power output, even when all other system parameters remain fixed.

\section{Alfvén wing structure and energy flux\label{sec:alf_wing_str_en_flx}}
\begin{figure}
    \centering
    \includegraphics[width=1.0\linewidth]{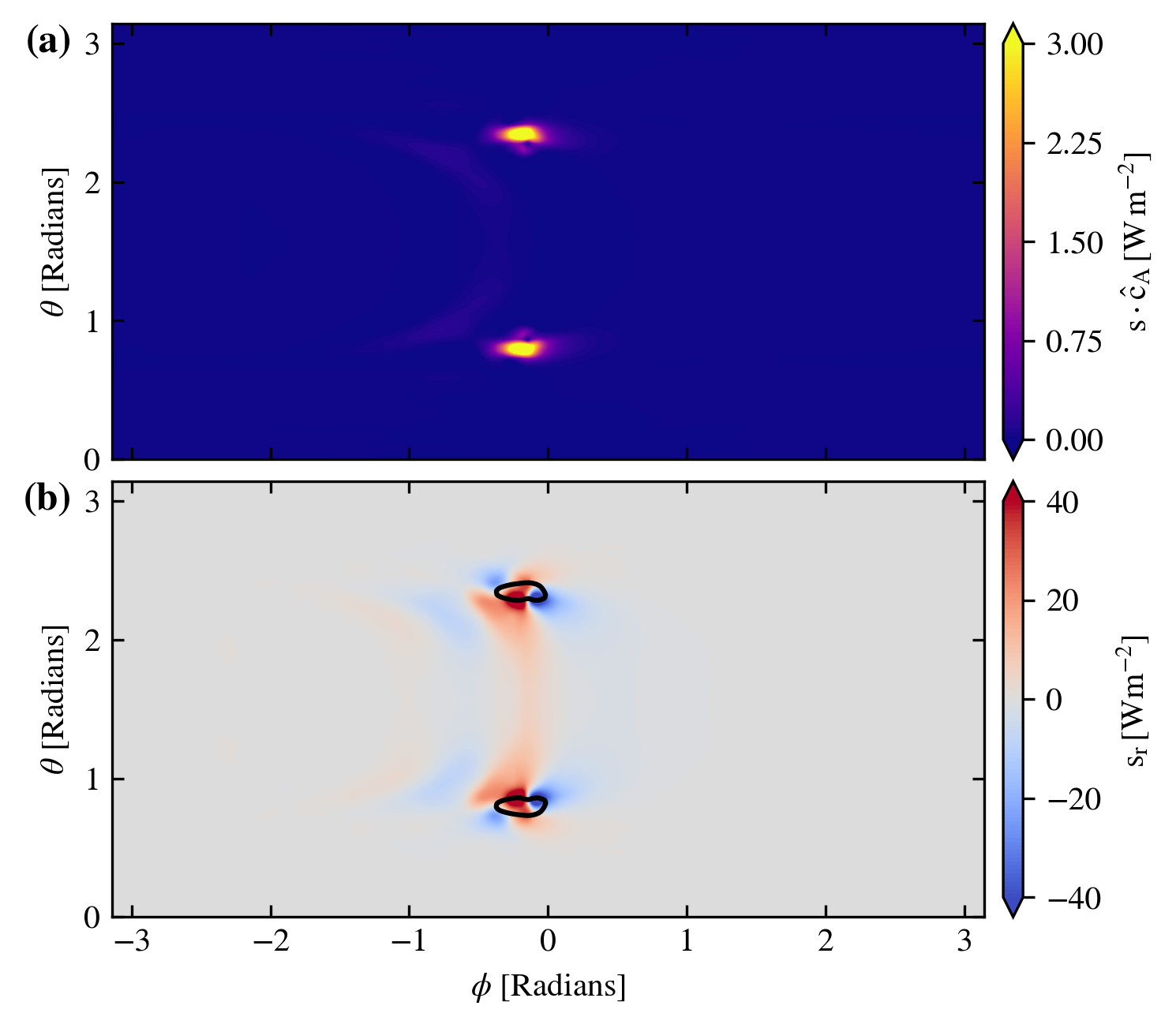}
    \caption{Panel (a) displays the quantity $\vec{s} \cdot \vec{\hat{c}}_A$ on a spherical surface of radius $2.5R_{\star}$ centered on the star. Panel (b) shows the radial component of the Poynting vector, $s_r$, on the same spherical surface. The black contours in this panel indicate lines of constant $(\vec{s} \cdot \vec{\hat{c}}_A) = 1\ \text{W m}^{-2}$.}
    \label{fig:Sr_andSca_2D}
\end{figure}

 Building on this context, we now proceed by comparing our simulation results to the Saur model first, which provides an analytical estimate of the Poynting flux directed along the system’s Alfvén wings. This flux is quantified by the scalar product $\vec{s} \cdot \vec{\hat{c}}_A$, which represents the projection of the Poynting vector $\vec{s}$ along the characteristic unit vector of $\vec{c_A}$. The quantity $\vec{c_A}$ corresponds to the Alfvén characteristics, defined as $\vec{c_{A{\pm}}} = \vec{v} \pm \vec{v_A}$, where $\vec{v}$ is the local plasma velocity. Figure~\ref{fig:Sr_andSca_2D} illustrates results for a configuration where the planet orbits at a radius of $5R_{\star}$. In panel (a), the spatial distribution of $(\vec{s} \cdot \vec{\hat{c}}_A)$ is shown on a spherical shell with radius $2.5R_{\star}$ centered on the star, similar to the integration surfaces depicted in Figure\ref{fig:3d_schematic}. The polar coordinate system used in figure \ref{fig:Sr_andSca_2D} represents the surface of a sphere centered on the star. By convention, the azimuthal angle $\phi = 0$ corresponds to the star–planet line, with $\phi$ ranging from $-\pi$ to $\pi$. The polar angle $\theta$ is defined such that $\theta = 0$ corresponds to the south stellar pole and increases to $\theta = \pi$ at the north stellar pole.
 
 As expected, the most intense regions of $(\vec{s} \cdot \vec{\hat{c}}_A)$ are concentrated at the intersections of the spherical surface with the Alfvén wings generated by the planet. This localization of enhanced Poynting flux directed along the Alfvén characteristics, aligns with previous studies and has been extensively discussed in the literature, confirming the role of Alfvén wings as primary channels for energy exchange in star–planet magnetic interactions.\citep{Strugarek_2014, Strugarek_2015, Strugarek_2016, Strugarek_2018}.

It is important to note that the quantity $(\vec{s} \cdot \vec{\hat{c}}_A)$ does not directly represent the total Poynting flux directed toward the star; rather, it quantifies the alignment of the Poynting vector $\vec{s}$ with the Alfvén characteristics $\vec{c_A}$. Physically, the flux actually directed toward the star is more accurately described by the negative components of the radial Poynting vector, $s_r$, evaluated on spherical shells centered on the star (as illustrated in Figure \ref{fig:3d_schematic}), where $(s_r, s_{\theta}, s_{\phi})$ denote the spherical components of $\vec{s}$. Panel (b) of Figure \ref{fig:Sr_andSca_2D} shows the corresponding spatial distribution of $s_r$ on the same surface as panel (a). Unlike the scalar product $(\vec{s} \cdot \vec{\hat{c}}_A)$, which is predominantly positive within the Alfvén wing and indicates directional alignment, the radial component $s_r$ takes on both positive and negative values. These correspond, respectively, to energy flowing outward from or inward toward the spherical surface. Additionally, the magnitudes of $s_r$ are substantially larger, as indicated by the differing color bar scales in both panels. In panel (b), black contours mark regions where $(\vec{s} \cdot \vec{\hat{c}}_A) = 1\ \text{W m}^{-2}$, delineating a reference cross section of the Alfvén wings in relation to the bipolar radial Poynting flux pattern. This cross section encompasses both inward and outward $s_r$ regions, implying that energy transport within the Alfvén wings is bi-directional, i.e., flowing both toward and away from the planet (or star). This behavior is consistent with local energy conservation, as the planetary obstacle primarily channels the available energy rather than being a source or sink. Consequently, any inward flux must be compensated by energy replenishment from the surrounding medium, predominantly the stellar wind, but also potentially supplemented by energy flow along the Alfvén wing magnetic field lines.

To explicitly estimate the power transferred toward the star, we focus on the inward-directed component of the radial Poynting flux. In the following sections, this is done by isolating the negative values of $s_r$, which correspond to the blue regions shown in panel (b) of Figure~\ref{fig:Sr_andSca_2D}.

\section{SPMI power estimates\label{sec:SPMIpowerestimates}}
To compare the power predicted by the two quantities $(\vec{s} \cdot \vec{\hat{c}}_A)$ and $s_r$, we compute surface integrals over the spherical shells centered on the star, as illustrated in Figure \ref{fig:3d_schematic} and labeled as the ``surfaces of integration''. These shells extend to radii just outside the planetary magnetospheric boundary (for example, up to $4.5 R_{\star}$ for a planet orbiting at $5 R_{\star}$). The resulting surface-integrated powers are then defined as
\begin{align}
S_r &= - \int \min(s_r, 0) \, r^2 \sin\theta \, d\theta \, d\phi \label{eq:sr_integral}\\  
S\cdot \hat{c}_A &= \int (\vec{s} \cdot \vec{\hat{c}}_A) r^2 sin\theta\, d\theta\, d\phi \label{eq:sca_integral}
\end{align}
Here, $\min(s_r, 0)$ selects only the negative values of the $s_r$ profile for integration, effectively excluding contributions from regions where the flux is directed outward. Consequently, the integral in Equation \ref{eq:sr_integral} captures only the power flowing inward toward the star and this integral therefore represents the absolute upper limit of the power made available in order to generate hotspots on the star. By convention, surface-integrated quantities are denoted with a capital S in units of watts, while the corresponding areal distributions use lowercase `s' in units of $\rm {Wm^{-2}}$.

 \begin{figure}
    \centering
    \includegraphics[width=1.0\linewidth]{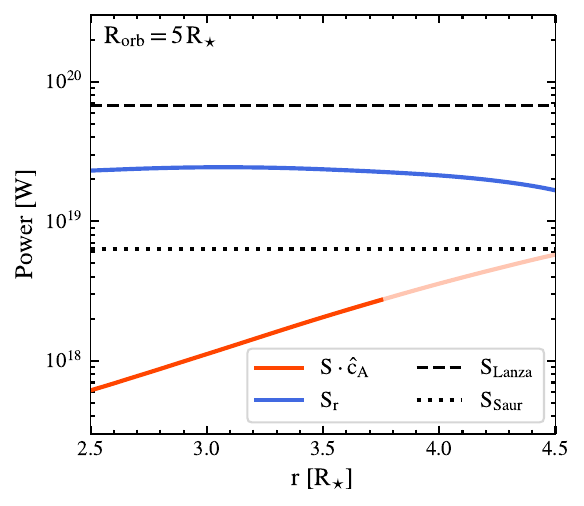}
    \caption{Surface-integrated quantities $S_r$ and $S \cdot \hat{c}_A$ shown as functions of radius for the system with an orbital radius of $5 R_{\star}$. The surface integration is performed over spherical shells of varying radii centered on the star, following the approach illustrated in figure \ref{fig:3d_schematic} and using the formulations in equations \ref{eq:sr_integral} and \ref{eq:sca_integral}. The translucent portion of the solid red ($S \cdot \hat{c}_A$) curve corresponds to integration radii where the chosen surfaces are suboptimal for evaluating the integral (see explanation in appendix \ref{sec:surface_int_truncated}). Consequently, the solid portion of the red curve highlights the range over which the integrated quantity is considered most reliable. The dotted and dashed black horizontal lines represent the corresponding analytical predictions from the Saur and Lanza models, as given by equations \ref{eq:Saur_power} and \ref{eq:Lanza_power}, respectively.}
    \label{fig:Sr_andSca_vs_r}
\end{figure}

Figure \ref{fig:Sr_andSca_vs_r} displays the surface-integrated quantities $S_r$ (blue line) and $S \cdot \hat{c}_A$ (red line) as functions of the radius of spherical shells centered on the star, for a planet situated at $R_{\text{orb}} = 5 R_{\star}$. The $S \cdot \hat{c}_A$ curve has been transitioned to a translucent version at a radius ($\sim 3.8 R_{\star}$), since beyond this point the spherical shells exclude parts of the $\vec{s}\cdot\vec{\hat{c}}_A$ structure, leading to a less robust estimate of the surface-integrated quantity. Accordingly, only the solid, opaque segment of the red curve represents a robust estimate of the integrated quantity. In contrast, the $S_r$ profile is not affected by this limitation, allowing surface integrals to be computed up to $\sim 4.5 R_{\star}$. A more detailed discussion of this distinction is provided in Appendix \ref{sec:surface_int_truncated}.

The decreasing trend of $S \cdot \hat{c}_A$ as one moves inward from the planet toward the star suggests that the Poynting vector becomes increasingly misaligned with the Alfvén characteristics near the star. However, this does not necessarily imply that the Poynting flux itself is no longer directed toward the star. The blue curve, representing the surface-integrated inward (negative) radial Poynting flux, exhibits a slight increase just outside the planet’s magnetosphere before flattening out closer to the star. This behavior indicates a mechanism that generates additional Poynting flux outside the magnetosphere, with the subsequent plateau reflecting energy conservation across spherical shells.

We explore the physical origin of this increase in inward-directed Poynting flux between the planet and the star in significant detail in Appendix \ref{sec:appendixA}. To summarize, our analysis reveals that all magnetic field lines forming the Alfvén wings, which connect the planet to the star, collectively act as an extended obstacle to the stellar wind flow. This contrasts with the prevailing view in the literature, which asserts that only the planet itself and its immediate surroundings (e.g., magnetosphere) act as obstacles, while the Alfvén wings are considered merely as channels conveying the power generated by that obstacle. Our results identify two key effects arising from this extended obstacle. First, the stellar wind is partially deflected toward the star across the obstacle’s full extent. Second, the interaction enhances the local magnetic field via flux pileup, resulting in an enhanced advection of magnetic energy toward the star. Together, these effects channel electromagnetic energy back toward the star, not necessarily confined to the Alfvén wings alone, but generally aligned along the same direction, leading to the observed increase in inward Poynting flux in regions where these interactions are significant.

For comparison, power predictions from the Saur and Lanza models, calculated by applying local simulation parameters to equations \ref{eq:Saur_power} and \ref{eq:Lanza_power}, are shown as dotted and dashed horizontal lines, respectively. An analysis across different orbital radii (plot not shown) reveals that the discrepancy between simulation results and analytical predictions varies with the planet’s orbital distance. At smaller radii (e.g., $R_{\text{orb}} = 3 R_{\star}$), the simulated $S_r$ slightly exceeds even the Lanza model’s prediction. At larger radii (e.g., $R_{\text{orb}} = 7 R_{\star}$), $S_r$ falls between the power levels predicted by the Saur and Lanza models. Overall, it is clear that neither analytical model fully captures the total power transported from the planet to the star as observed in our simulations.

Previous validations of the Saur model have shown reasonable agreement between observed, predicted, and simulated values of $S \cdot \hat{c}_A$ integrated over the Alfvén wing cross section \citep{Strugarek_2016}, consistent with the model’s goal of estimating power generated at the planet and carried within the Alfvén wings. This agreement is also evident in the simulations presented here as seen from figure \ref{fig:Sr_andSca_vs_r}, when focusing on $S \cdot \hat{c}_A$. In particular, in the regions near the planet, as indicated along the right side of the red curve, the $S \cdot \hat{c}_A$ profile shows strong agreement with the Saur model prediction. This consistency is further examined in Appendix \ref{sec:surface_int_truncated}, where surface integrations centered on the planet are specifically considered for this case, in order to minimize uncertainties associated with the choice of integration surfaces. However, it is important to emphasize that alignment between the Poynting vector $\vec{s}$ and the unit Alfvén characteristic $\vec{\hat{c}}_A$ does not necessarily imply net energy transport toward the star. As illustrated in panel (b) of Figure~\ref{fig:Sr_andSca_2D}, the Alfvén wing cross section inherently contains regions of bidirectional energy flow, with components directed both toward and away from the star. Furthermore, our results reveal significant excess power generation immediately outside the planet’s magnetosphere. This additional energy boosts the radial Poynting flux component ($S_r$), contributing power toward the star that lies beyond the original scope of the Saur model. Additionally, as will be discussed in Section \ref{sec:Jovian_analogues_vs_spmi}, the Jovian system may not serve as a perfect analog for SPMI systems in all cases, highlighting the need for caution when drawing parallels between them.

\subsection{Dependence on planetary magnetic field strength}
\begin{figure}
    \centering
    \includegraphics[width=1.0\linewidth]{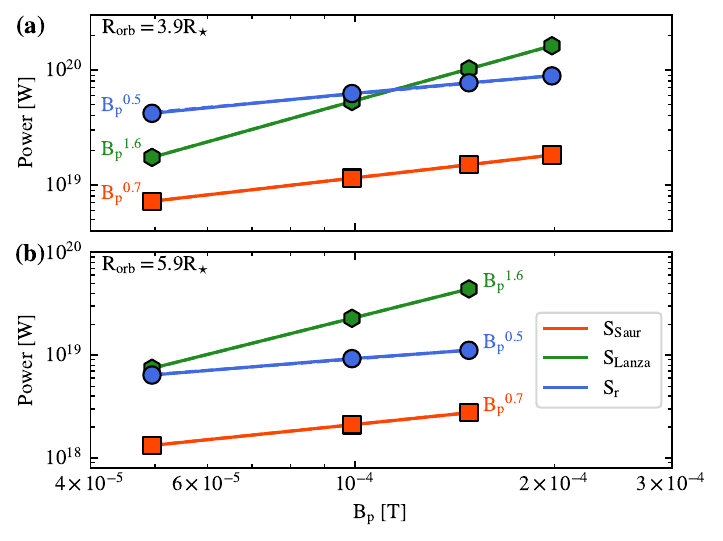}
    \caption{Comparison of SPMI power as a function of planetary magnetic field strength at two orbital radii. Panel (a) shows the results for $R_{\text{orb}} = 3.9,R_{\star}$, and panel (b) for $R_{\text{orb}} = 5.9,R_{\star}$. The scatter points indicate simulation results; the solid lines represent power-law fits. The blue markers and lines denote $S_r$ from simulations, the red squares and lines show the predictions from the Saur model, and the green hexagons and lines correspond to the Lanza model.}
    \label{fig:S_vs_Bp}
\end{figure}

Next, we examine how the Poynting flux varies with changes in the planetary magnetic field strength, $B_p$. For planets at orbital radii of 3.9 $R_{\star}$ and 5.9 $R_{\star}$, we conducted parameter studies varying only $B_p$ while keeping all other parameters constant. At 3.9 $R_{\star}$, four planetary surface magnetic field strengths were simulated, ranging from $5 \times 10^{-5}$ T to $2 \times 10^{-4}$ T (0.5 G to 2 G). At 5.9 $R_{\star}$, three values were tested, spanning from $5 \times 10^{-5}$ T to $1.5 \times 10^{-4}$ T (0.5 G to 1.5 G). Figure \ref{fig:S_vs_Bp} presents the variation of $S_r$, $S_{\text{Saur}}$, and $S_{\text{Lanza}}$, computed using equations \ref{eq:sr_integral}, \ref{eq:Saur_power}, and \ref{eq:Lanza_power}, respectively, as functions of the planetary equatorial magnetic field strength.

In figure \ref{fig:S_vs_Bp}, panel (a) shows results for an orbital radius of 3.9 $R_{\star}$, while panel (b) corresponds to 5.9 $R_{\star}$. Scatter points represent simulation data, and solid lines are power-law fits. Blue points and lines correspond to $S_r$. Consistent with previous analyses, $S_r$ values are roughly an order of magnitude higher than those predicted by the Saur model (red squares and line). While the slopes of $S_r$ and the Saur model are similar, their magnitudes differ significantly. Conversely, the magnitude of $S_r$ is closer to the Lanza model predictions (green hexagons and line), though the power-law index differs markedly. Specifically, the simulations yield a scaling of $S_r \propto B_p^{0.5}$, compared to $S_{\text{Saur}} \propto B_p^{0.7}$ and the steeper $S_{\text{Lanza}} \propto B_p^{1.6}$ over the range of planetary magnetic field strengths considered.

\subsection{A new scaling law for SPMI power}
Based on the analysis above, it is evident that existing scaling laws do not fully capture the power transferred toward the star. Therefore, we propose a new scaling law, grounded in the results of our numerical simulations. We reiterate here that varying the planet’s orbital radius simultaneously influences several key aspects of the SPMI system. Most notably, it changes the planet’s orbital velocity, which affects its relative velocity with respect to the ambient stellar wind plasma ($v_{\text{rel}}$). At the same time, it alters the local stellar magnetic field strength at the planet’s position ($B_W$). Both these parameters play a critical role in determining the total power available at the planet’s location, only a fraction of which is eventually channeled back toward the star. In practice, decoupling these two quantities, $v_{\text{rel}}$ and $B_W$, within our current simulation set is challenging, since modifying the orbital radius inherently changes both parameters in any given stellar wind model simultaneously. However, we find that an explicit decoupling may not be strictly necessary. By normalizing the obtained power $S_r$ with a characteristic Poynting flux scale $(\pi R_p^2 B_W^2 v_{\text{rel}} / \mu_0)$, the resulting normalized power reveals a clear modulation on the planetary and stellar magnetic fields. This relationship can be expressed in the form:

\begin{equation}\label{eq:scaling_law}
\rm{\frac{S_{r} \mu_0}{(\pi R_p^2\,B_W^2\,v_{rel})} = \mathcal{A}\, B_p^\alpha \,B_W^\beta}
\end{equation}
where the parameters $\mathcal{A}$, $\alpha$, and $\beta$ are to be determined by fitting the simulation results. The normalizing factor $(\pi R_p^2 B_W^2 v_{\text{rel}} / \mu_0)$, carries an important physical distinction. It corresponds to the characteristic power intercepted by the cross section of a planetary obstacle whose magnetosphere is just large enough to match the size of the planet body at the orbital location of the planet. The right-hand side then captures how this intercepted power is modulated by variations in the planetary magnetic field strength $B_p$ and the local stellar wind magnetic field $B_W$. 

 \begin{figure}
    \centering
    \includegraphics[width=1.0\linewidth]{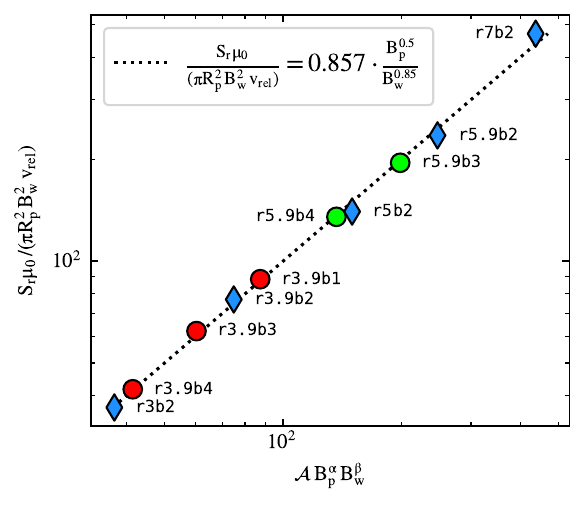}
    \caption{New scaling law for the power directed toward the star via SPMIs, as described by equation \ref{eq:scaling_law}. The data points are drawn from all simulation setups listed in Table \ref{tab:simulation_setups}. The blue diamonds indicate the default cases, while the red and green circles represent cases with varying planetary magnetic field strengths. For clarity, each point is labeled with the corresponding setup name from Table \ref{tab:simulation_setups}.}
    \label{fig:Scaling_fit}
\end{figure}

Figure \ref{fig:Scaling_fit} illustrates the proposed scaling law. As shown, the variations listed in Table \ref{tab:simulation_setups}, including changes in $v_{\text{rel}}$, $B_W$, and $B_p$, collapse cleanly onto the trend predicted by equation \ref{eq:scaling_law}. The resulting fit is remarkably consistent across all the cases simulated, providing strong support for the robustness and general applicability of the proposed relation. Guided by the fit in Figure \ref{fig:Scaling_fit}, we express the dimensional form of our scaling law as

\begin{equation}\label{eq:final_scaling_law}
\rm{S_{r} = 0.857\times \left( \frac{\pi R_p^2 B_w^2 v_{rel}}{\mu_0}\right) \left[ \frac{B_p}{1 T}\right]^{0.5} \left[ \frac{1T}{B_w}\right]^{0.85} \:\:\: [Watts].}
\end{equation}

This expression estimates the total power transferred toward the star due to sub-Alfvénic star–planet magnetic interaction (SPMI), where $R_p$ is the planetary radius, $B_W$ denotes the local stellar wind magnetic field strength at the planet’s orbit, $v_{\text{rel}}$ is the relative velocity between the stellar wind and the planet, and $B_p$ represents the planet’s equatorial surface magnetic field strength. All quantities in the scaling law are given in SI units. This scaling law applies to the maximum power transferred from the planet to the star, which occurs in the case of an open magnetosphere. The closed magnetospheric scenario is examined in Section \ref{sec:closed_magnetosphere_scaling}. Importantly, since this relation is derived from values computed on concentric spherical shells centered on the star and given that $S_r$ stabilizes close to the stellar surface, the expression in equation \ref{eq:final_scaling_law} provides a reliable estimate of the SPMI power reaching just outside the star’s transition region. This is particularly significant, as it implies that the scaling law inherently accounts for the physical processes and dissipation losses occurring along the path from the planet to the star, at least within an MHD approximation, up to the base of the stellar corona. To evaluate the quality of the scaling law fit in comparison to the simulation results, we quantify the residuals of the fit by using the mean absolute relative deviation (MARD), defined as
\begin{equation}
\mathrm{MARD} = \frac{1}{n} \sum_{i=1}^{n} \left| \frac{Y_{\mathrm{sim},i} - Y_{\mathrm{scaling},i}}{Y_{\mathrm{sim},i}} \right| \times 100 \;\;\;[\%]\label{eq:MARD}
\end{equation}
where $Y_{\mathrm{sim},i}$ are the values obtained from the simulations, and $Y_{\mathrm{scaling},i}$ are the corresponding values predicted by the scaling law; \textit{n} represents the total number of data points, which in this case is ten. The MARD value for the fit of equation \ref{eq:final_scaling_law} is $\sim 3\%$, indicating an excellent agreement.

\subsection{Dependence on planetary magnetic field topology\label{sec:closed_magnetosphere_scaling}}
It is well established that the power a planet can channel toward its host star is strongly influenced by its magnetic topology, particularly the orientation of its dipole moment relative to the stellar magnetic field. This alignment gives rise to two limiting configurations: the open and closed magnetospheres (e.g., Figure 2 in \citealt{Ip_2004}). An open magnetosphere forms when the planetary and stellar magnetic field lines are oppositely directed near the planet, promoting efficient magnetic reconnection and strong star-planet interaction. In contrast, a closed magnetosphere emerges when the fields are aligned, resulting in a markedly different local magnetic topology that reduces the net energy transfer. The effects of these configurations have been thoroughly explored in prior work \citep{Strugarek_2014, Strugarek_2015}.

At present, tentative observational methods used to probe planetary magnetic fields by leveraging SPMI, lack the precision needed to resolve differences in SPMI power associated with various dipole orientations \citep{Cauley_2018, Cauley_2019}. Therefore, conducting a detailed parameter scan over dipole tilt angles may not provide observationally actionable insights at this time. Nonetheless, it remains essential to establish a conservative lower bound on the SPMI power that a magnetized planet can produce. This lower limit is effectively represented by the closed magnetosphere configuration.

As part of our simulation suite, we explored the closed magnetosphere configuration at three different orbital radii: $3.9 R_{\star}$, $5.9 R_{\star}$, and $7 R_{\star}$. We find that the SPMI power ($S_r$) obtained in these cases fits relatively well the same functional form as the scaling law presented in equation~\ref{eq:final_scaling_law}, with the only difference being in the value of the prefactor $\mathcal{A}$. Specifically, for the closed magnetosphere configuration, the prefactor takes the value $\mathcal{A} = 0.169$. The corresponding scaling relation for the closed case can thus be written as

\begin{equation}\label{eq:closedMS_scaling_law}
\rm{S_{r\, (closed\, MS)} \simeq \frac{S_{r\, (open \,MS)}}{5}},
\end{equation}

where $\rm{S_{r\, (open\, MS)}}$ corresponds to $S_r$ in equation \ref{eq:final_scaling_law}, originally representing the open magnetosphere case. This indicates that the SPMI power in the closed magnetosphere configuration is approximately one-fifth of that in the open magnetosphere scenario. This metric exhibits a comparatively lower quality of fit to the simulation data, with a MARD of approximately 30\% as calculated using equation \ref{eq:MARD} based on a limited sample of three data points with closed magnetospheric configurations. However, since the primary focus of this paper is to estimate the absolute maximum power (upper limit) that a planet can channel toward its host star, the relatively larger deviation in fitting the lower limit can be regarded as acceptable within the scope of this work. Therefore, equation \ref{eq:closedMS_scaling_law} should be viewed as offering a guiding estimate for the lower bound (i.e., the closed magnetospheric scenario), rather than representing a strict or absolute limit.

\section{Discussions and conclusions}
\label{sec:disc_and_conc}

\subsection{Jovian analogs vs SPMIs\label{sec:Jovian_analogues_vs_spmi}}

\begin{figure}
    \centering
    \includegraphics[width=1.0\linewidth]{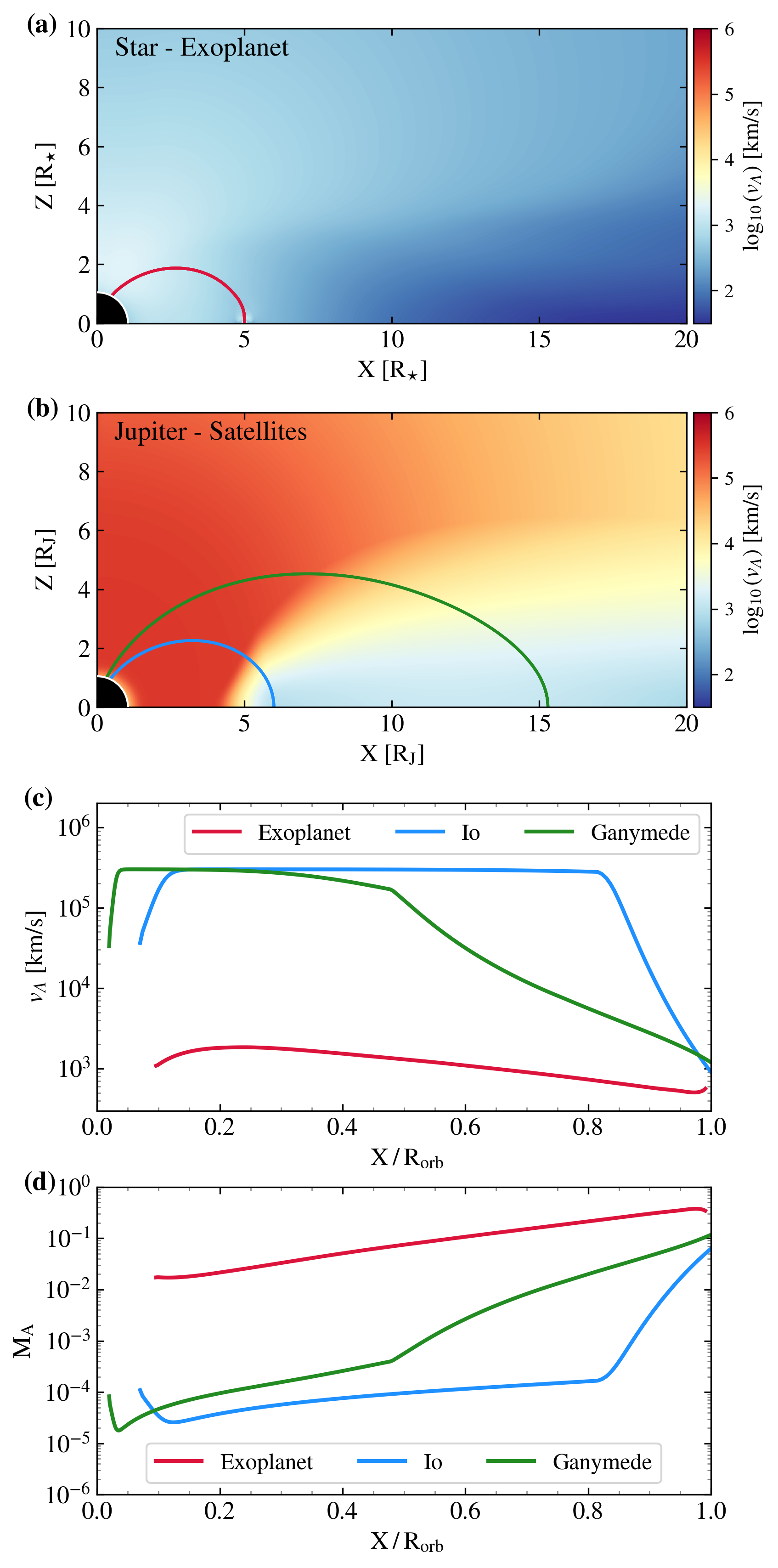}
    \caption{Comparison of Alfvén speed and Alfvénic Mach number profiles in SPMI and the Jovian system. Panel (a) shows the Alfvén speed distribution in our simulated SPMI system. The red curve traces a magnetic field line from an exoplanet at 5 stellar radii to the star. Panel (b) presents a meridional slice of Alfvén speed within Jupiter’s magnetosphere, highlighting the localized influence of the Io plasma torus. The blue and green lines mark magnetic field lines traced from the Io and Ganymede orbits, respectively. Panel (c) plots the Alfvén speed along these field lines and  panel (d) shows the corresponding Alfvénic Mach number profiles along the field lines.}
    \label{fig:Jupiter_vs_spi}
\end{figure}

It is clear that although the Saur model performs reasonably well for sub-Alfvénic interactions in the Jovian magnetosphere \citep{Saur_2013,Kotsiaros_2024}, it significantly underestimates the SPMI power predicted by the simulations. This discrepancy raises important questions about which Jovian analogs can be confidently extended to explain SPMI observations and which should be applied with restraint. In the following paragraphs, we examine why a model effective for sub-Alfvénic cases in the Jovian magnetosphere does not accurately capture sub-Alfvénic star--planet interactions.

As an important distinction, the analytical expression derived by the Saur model given by equation \ref{eq:Saur_power} has been simplified under the approximation that $M_A \rightarrow 0$. We therefore examine the Alfvén speed profile of the Jovian magnetosphere and compare it with those in SPMI environments. In the case of SPMI, the stellar wind is generally expected to be relatively homogeneous in density on concentric spheres, and due to the smooth variation of the stellar magnetic field, the Alfvén speed profile typically varies by only about 4–5 times between the planet and the star. While this variation may be somewhat enhanced by localized inhomogeneities around particularly active stars, such features are typically spatially confined, so the environment remains relatively uniform overall. Panel (a) of figure \ref{fig:Jupiter_vs_spi} shows the Alfvén speed profile within the simulated SPMI system. The red curve highlights a magnetic field line tethering the star to the close-in exoplanet at an orbital radius of $r_{\rm orb} = 5R_{\star}$. The 2D profile indeed appears quite homogeneous between the star and the planet. Note that the colorbar limits are relatively high in panel (a) and have been kept so in order to be consistent between this plot and the subsequent one explained in the paragraph below.

The Jovian magnetosphere, however, is more complex. While planetary inner magnetospheres are generally expected to be sparsely populated with low plasma density, surface activity of Io, orbiting at about 6 $R_J$, locally enhances the density within Jupiter’s magnetosphere. This localized increase in plasma density then radially diffuses outward to larger orbital radii, creating a doughnut-shaped region known as the Io plasma torus. Additionally, the Io plasma torus harbors currents that locally influence Jupiter’s large-scale magnetic field, leading to radial and meridional anisotropies in both plasma density and magnetic field profiles. As a result, the Alfvén speed distribution within Jupiter’s magnetosphere is highly variable. Panel (b) of figure \ref{fig:Jupiter_vs_spi} shows a meridional slice of the Alfvén speed inside Jupiter’s magnetosphere. The plasma density within the torus is modeled following \citet{Lysak_2020,Su_2006} and \citet{Bagenal_2011} with a floor value of 0.01${\rm cm}^{-3}$, while the magnetic field is taken as a superposition of Jupiter’s dipole field and the residual fields produced by the plasma torus, as prescribed by \citet{Connerney_1981} with updated corrections of \citet{Connerney_2020}. Such a slice of Alfvén speed profile was first presented by \citet{Lysak_2020} (figure 1(c) in \citet{Lysak_2020}) and has been reproduced here using similar parameters and relativistic corrections \citep{Su_2006} to enable qualitative comparison with the current work.

As seen in panel (b) of figure \ref{fig:Jupiter_vs_spi}, the Alfvén speed profile within this meridional slice is indeed highly variable. The blue and green lines in panel (b) represent magnetic field lines traced from the orbital locations of Io and Ganymede, at $6R_J$ and $15.3R_J$, respectively. To enable a relative comparison, the colorbars in panels (a) and (b) of figure \ref{fig:Jupiter_vs_spi} have been set to the same scale. Panel (c) of figure \ref{fig:Jupiter_vs_spi} shows the Alfvén speed along the magnetic field lines depicted in panels (a) and (b), plotted in their corresponding colors. The abscissa in panel (c) is normalized to the orbital radius of the respective orbiting body. Near the orbital position ($X/R_{\rm orb} \sim 1$), the Alfvén speed is comparable between the Jovian and exoplanetary cases. However, as one moves closer to the central object, the Alfvén speed along the magnetic field lines traced from the Jovian satellites diverges significantly from the SPMI case, with differences reaching up to about three orders of magnitude. 

As an approximation, we consider Alfvén wings to act as rigid rotators within the stellar wind, with an angular speed determined by the relative velocity of the orbiting body with respect to the surrounding plasma. Based on this assumption, and using the known relative speed of the body, we compute the Alfvénic Mach number along each segment of the magnetic field line connecting the orbiting body to its host. For this analysis, we adopt a relative speed of 195 km s$^{-1}$ for the exoplanet, corresponding to its orbital velocity. The relative plasma velocity for Io is taken to be 57 km s$^{-1}$ \citep{Futaana_2015}, and for Ganymede, 140 km s$^{-1}$ \citep{Kivelson_2004}. Using this radial profile of the relative speed of the Alfvén wings with the surrounding plasma, we compute the local Alfvénic Mach number along each infinitesimal segment of the field lines. The resulting profiles are shown in panel (d) of figure \ref{fig:Jupiter_vs_spi} with the abscissa similar to panel (c). Near the orbit of the exoplanet, the Alfvénic Mach number ($M_A$) is approximately 0.4, while for the Jovian moons it is lower and ranges between $10^{-2}$ to just above $10^{-1}$ near the orbiting body. This indicates that the Jovian moons already operate in a regime more consistent with the low-$M_A$ approximation used in the derivation of the analytical expression in the Saur model (i.e., where $M_A \rightarrow 0$). This approximation becomes increasingly valid along the magnetic field lines moving inward toward Jupiter, where $M_A$ drops to values as low as $10^{-4}$ to $10^{-5}$.

A key assumption of the Saur model is that the SPMI power propagating along the Alfvén wings is primarily generated in close proximity to the planetary obstacle. In the context of the Jovian moons, where the Alfvénic Mach number, $M_A$, attains relatively elevated values only near the moons and decreases sharply toward Jupiter, the Saur model’s proportionality of power generation to $M_A$ at the obstacle, where orbital velocity dominates, justifies the expectation that the majority of the interaction power originates near the moons. In contrast, for SPMI systems, panel (d) of figure \ref{fig:Jupiter_vs_spi} demonstrates that although $M_A$ decreases along the magnetic field lines from the planet toward the star, the decline remains noticeably moderate. The minimum $M_A$ values obtained for SPMI remain comparable to those at the Jovian moon orbital distances. Consequently, in the SPMI scenario, the entire Alfvén wing can act as an extended obstacle, with distributed contributions to the total Poynting flux directed toward the host star. This distributed power generation mechanism is consistent with SPMI simulation results, which reveal a progressive increase in inward propagating power with distance from the planet. We attribute these two factors as the primary reasons why the Saur model underestimates the power in SPMI systems, while still providing reasonably accurate approximations for the Jovian case. We also note here that the Lanza model of SPMI power primarily emphasizes a stretch-and-break mechanism of magnetic field lines, which facilitates the transfer of magnetic stresses to the star due to the planet’s orbital motion. This particular mechanism is not straightforward to probe directly within the scope of our current simulations, making it challenging to verify or interpret based on the present results. For this reason, we have chosen not to compare our interpretation with the Lanza model’s mechanism here, but we believe this offers an interesting avenue for future investigation.  
\subsection{Practical applications of the new scaling law}
 \begin{figure}
    \centering
    \includegraphics[width=1.0\linewidth]{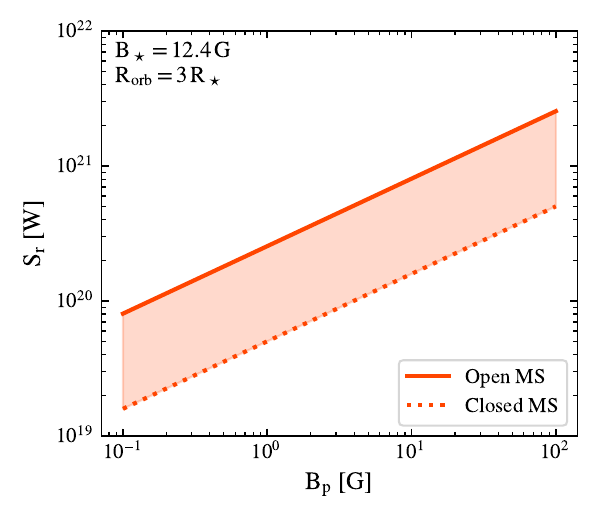}
    \caption{Estimated power directed toward the star by a Jupiter-sized planet as a function of planetary magnetic field strength, orbiting at a distance of $3 R_{\star}$ around a star with a dipolar equatorial magnetic field of 12.4 G. The solid line indicates the upper bound for an open magnetosphere scenario, while the dotted line corresponds to a closed magnetosphere case.}
    \label{fig:Scaling_law_prediction}
\end{figure}
We now consider the practical application of the scaling law developed herein. In our simulations, the stellar equatorial surface magnetic field is assumed to be comparable to that of the Sun within an order of magnitude, approximately 12.4 G, assuming a dipolar magnetic topology. The background stellar wind model directly provides the magnetic field strength, $B_W$, at the planetary orbital radius. In the absence of detailed stellar wind simulations, $B_W$ can be reasonably approximated by applying a radial decay law, for example, $B_W \propto r^{-a}$, provided the stellar surface field $B_{\star}$ is known from Zeeman Doppler Imaging (ZDI) measurements. The scaling law also requires the relative velocity, $v_{rel}$, between the planet and the ambient stellar wind plasma. Within our wind model framework, $v_{rel}$ is computed as the vector difference of the planetary orbital velocity and the local stellar wind velocity. Alternatively, the orbital velocity, which can be easily obtained from observations of exoplanetary systems, may serve as a reasonable approximation for $v_{rel}$ in close-in systems around slowly rotating stars or systems where the stellar rotation period is significantly larger than the planetary orbital period. The planetary radius, $R_p$, is set to 0.1 $R_{\star}$ in our simulations, roughly corresponding to a Jupiter-sized planet orbiting a Sun-like star. Observationally, $R_p$ can be reliably inferred from transit measurements. With these parameters specified in SI units, equation \ref{eq:final_scaling_law} can be directly employed to estimate SPMI power across a range of planetary magnetic field strengths. Figure \ref{fig:Scaling_law_prediction} presents the power directed toward the star as a function of planetary magnetic field magnitude at an orbital radius of $3 R_{\star}$ for a Sun like stellar magnetic field. The solid curve represents the upper limit for an open magnetosphere scenario (equation \ref{eq:final_scaling_law}), while the dotted curve corresponds to a closed magnetosphere scenario (equation \ref{eq:closedMS_scaling_law}). Importantly, the scaling law predicts that planetary magnetic fields of hot-Jupiters on the order of 10-100 G can generate SPMI power exceeding $10^{21}$ W, with power output increasing further for stronger stellar magnetic fields, highlighting the potential significance of SPMI energetics in such systems.

In conclusion, this paper proposes a new scaling law (Eq. \ref{eq:final_scaling_law}) to estimate the power transported from a planet to its host star via star-planet magnetic interactions (SPMIs). This law is derived from comprehensive numerical simulations that reveal limitations in existing analytical models. Our analysis demonstrates systematically larger upper limits for the power values, up to an order of magnitude higher, along with somewhat different scaling exponents compared to previous studies by \citet{Saur_2013}, \citet{Lanza_2013}, and \citet{Strugarek_2016}. Table \ref{tab:comparison} summarizes the scaling of the normalized SPMI power $S$, where the normalization factor $v_{\rm rel} B^2$ is a scaling associated with the Poynting flux density from the stellar wind impacting the planetary obstacle. The table compares the exponents of $v_{\rm rel}$, $B_p$, and $B_w$ from two analytical models \citep{Saur_2013, Lanza_2013} and the most recent numerical study \citep{Strugarek_2016} of an SPMI scaling law. Focusing on the numerical results of \citet{Strugarek_2016}, their analysis finds the power scaling as $\propto B_p^{0.46}$ in the open magnetosphere case, which closely aligns with our scaling of $\propto B_p^{0.5}$. However, the dependence on the stellar wind magnetic field $B_w$ differs markedly. While \citet{Strugarek_2016} report that the normalized redirected power, $S_r / (B_w^2 v_{\rm rel})$, scales steeply as $\propto B_w^{-2.55}$, our results indicate a significantly shallower slope of approximately $B_w^{-0.85}$. As seen in Table \ref{tab:comparison}, the signs of the exponents are consistent across all models, reflecting agreement in the overall trends. However, the absolute values of the exponents highlight differences in the sensitivity of the power to each parameter. Notably, our analysis suggests that the entire Alfvén wing acts as an obstacle to the stellar wind, rather than just the immediate vicinity of the planet. This conceptual shift is reflected in the substantially weaker dependence on $B_w$ in our new scaling law, thereby producing comparatively larger Poynting fluxes even when considering weaker stellar wind magnetic field, $B_w$.
\begin{table}[]
\centering
\begin{tabular}{lcccc}
\toprule\toprule
                     & Saur & Lanza & Strugarek & This paper \\ 
\midrule\midrule

$\frac{S}{v_{\text{rel}} B_w^2}\propto$ & $B_{\text{p}}^{0.66}$ & $B_{\text{p}}^{1.66}$ & $B_{\text{p}}^{0.46}$ & $B_{\text{p}}^{0.5}$ \\
\midrule
$\frac{S}{v_{\text{rel}} B_w^2}\propto$ & $B_{\text{w}}^{-1.66}$ & $B_{\text{w}}^{-1.66}$ & $B_{\text{w}}^{-2.55}$ & $B_{\text{w}}^{-0.85}$ \\
\bottomrule
\end{tabular}
\caption{Comparison of key exponents from different studies.}
\label{tab:comparison}
\end{table}

Our formulation effectively captures the influence of key parameters such as orbital radius, planetary magnetic field strength, and stellar wind conditions, while also distinguishing between open and closed magnetospheric configurations. Importantly, this scaling law naturally incorporates the inefficiencies associated with power propagation toward the star and provides a useful tool for interpreting observational data, including the estimation of planetary magnetic fields from tentative detections of stellar hotspots linked to SPMIs. Nevertheless, it remains necessary to explicitly consider the inefficiencies caused by the partial reflection of Alfvén waves at the stellar transition region, as emphasized by \cite{Paul_2025}, when calculating the final energy deposition at the stellar surface. We also acknowledge that further work is needed, and is currently underway, to investigate the potential role of a planetary ionosphere on this scaling law. Nonetheless, this work represents an important step forward in advancing our understanding of the complex dynamics involved in star-planet magnetic interactions, and also serves as a crucial reconciliation of scaling laws inspired by the Jovian system, proposing their potential revisions to account for key differences and to ensure broader applicability to close-in exoplanets.

\begin{acknowledgements}
A.P. and A.S. express their sincere gratitude to Dr. Joachim Saur for his valuable insights. They also thank Dr. Philippe Zarka for drawing their attention to a relativistic correction which was originally omitted in the manuscript. Support from the MERAC Foundation is gratefully acknowledged. A.P. further thanks Dr. Robert Lysak for his prompt and helpful responses, which greatly accelerated the progress of this work. A.S acknowledges funding from the European Union’s Horizon 2020 research and innovation programme (grant agreement no. 776403 ExoplANETS-A), the PLATO/CNES grant at CEA/IRFU/DAp, and the European Research Council project ExoMagnets (grant agreement no. 101125367). Computational and storage resources for this project were provided by GENCI at TGCC under grant 2024-A0160410133, utilizing the Irene supercomputer’s Skylake and Rome partitions. Finally, the authors would like to thank the reviewer for their constructive comments, which have helped improve the quality and clarity of the manuscript.
\end{acknowledgements}
\bibliography{biblio}{} 
\bibliographystyle{aa}

\begin{appendix}
\FloatBarrier
\section{Surfaces of integration for $s_r$ and $\vec{s}\cdot\vec{\hat{c}}_A$ \label{sec:surface_int_truncated}}
\begin{figure}
    \centering
    \includegraphics[width=1.0\linewidth]{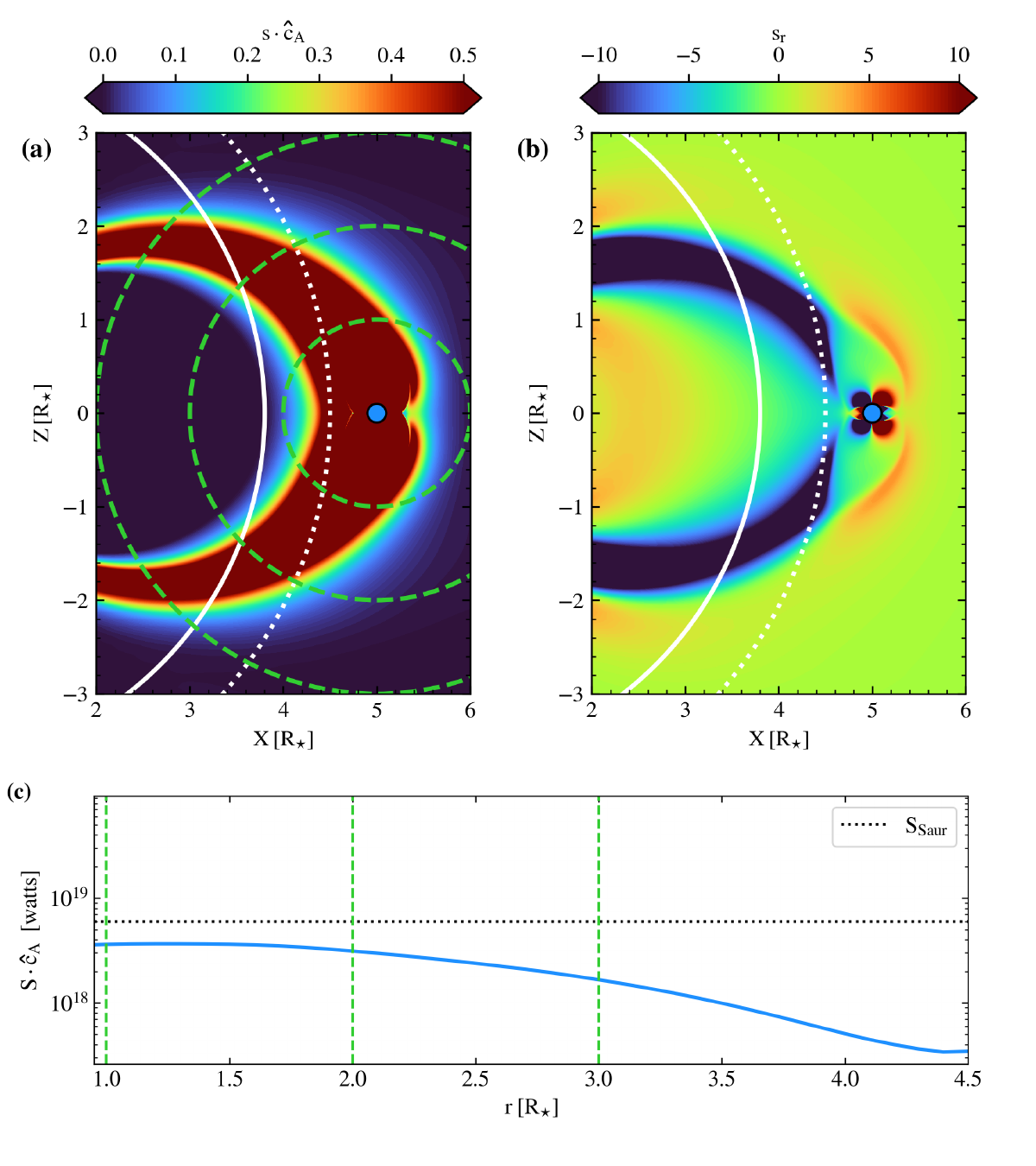}
    \caption{Panel (a) presents the meridional distribution of $\vec{s} \cdot \vec{\hat{c}}_A$, while panel (b) displays the $s_r$ distribution along the same cross-section. The solid and dotted white arcs indicate slices through spherical surfaces centered on the star, with radii of $3.8 R_{\star}$ and $4.5 R_{\star}$, respectively. In panel (a), the green dashed circles represent slices of spherical surfaces centered on the planet, used for alternative surface integrations. Panel (c) shows the surface integrated quantity $S \cdot \hat{c}_A$ integrated over various spherical radii centered at the planet. }
    \label{fig:scasr_circles}
\end{figure}

This section begins by explaining why the calculation of $S \cdot \hat{c}_A$ surface integrals is considered unreliable beyond a radius of approximately $3.8 R_{\star}$, as indicated by the transition from the opaque to the translucent curve in figure \ref{fig:Sr_andSca_vs_r}, while the surface integrals for $S_r$ can be reliably computed up to about $4.5 R_{\star}$. As shown in panel (a) of figure \ref{fig:scasr_circles}, the profile of $\vec{s} \cdot \vec{\hat{c}}_A$ is considerably broader (at Z = 0) than the negative portion of $s_r$ depicted in panel (b). The solid white arcs in both panels represent a slice of the sphere at a radius of $3.8 R_{\star}$. Within this radius, the spherical surface fully intersects the region containing the $\vec{s} \cdot \vec{\hat{c}}_A$ flux, ensuring that the surface integral accounts for nearly the entire flux crossing the sphere.

However, when we consider larger spheres, such as the dotted white arc in panel (a), which corresponds to a radius of $4.5 R_{\star}$, a significant part of the $\vec{s} \cdot \vec{\hat{c}}_A$ profile lies entirely inside the spherical surface of integration, rather than crossing its surface. This means the surface integral will fail to account for flux contained wholly within the volume, leading to an inaccurate estimation of the total flux crossing the sphere. Because surface integrals must ensure that all regions contributing to the flux cross the surface boundary, surfaces beyond $3.8 R_{\star}$ are considered to provide less reliable estimates of $S \cdot \hat{c}_A$. Therefore, in Figure \ref{fig:Sr_andSca_vs_r}, the solid opaque portion of the $S \cdot \hat{c}_A$ curve is limited to a maximum radius of $3.8 R_{\star}$ for the surface of integration. The color maps in panels (a) and (b) of figure \ref{fig:scasr_circles} have been intentionally saturated to clearly show the full spatial extent of the contributing regions. In contrast, the $s_r$ profile does not experience this limitation. Panel (b) of figure \ref{fig:scasr_circles} shows that at $3.8 R_{\star}$, the solid white arc fully intersects the negative $s_r$ regions without enclosing any flux-contributing areas entirely within the sphere. When the radius is increased to $4.5 R_{\star}$, represented by the dotted arc, the spherical surface continues to intersect all relevant negative portions of $s_r$ correctly, ensuring that no contributing region lies completely inside the spherical surface as was seen in panel (a). Consequently, the surface integral remains valid and meaningful, accurately capturing the total flux up to this larger radius. Additionally, to verify the agreement between the quantity $S \cdot \hat{c}_A$ obtained from the simulations and the total power estimate obtained from the Saur model, for the specific case of the planet at an orbital distance of $5 R_{\star}$, we compute surface integrals of $S \cdot \hat{c}_A$ over spherical surfaces centered at the planet instead of the star. The typical planet-centered spherical surfaces used for integration are shown as green dashed arcs/circles in the meridional slice of panel (a). Panel (c) illustrates the variation of the surface-integrated quantity $S \cdot \hat{c}_A$ for different radii of these planet-centered surfaces of integration. The Saur model estimate agrees very well with the surface-integrated $S \cdot \hat{c}_A$ values close to the planet (at smaller integration radii) and remains within a factor on the order of unity. To provide a reference of the values at specific integration surfaces, the radii of the three green dashed circles or arcs shown in panel (a) are also marked as green vertical lines in panel (c). Finally, it is important to note that performing surface integrals over spheres centered at the planet is only appropriate for the quantity $\vec{s} \cdot \vec{\hat{c}}_A$, consistent with the assumptions underlying the Saur model. For other quantities, such as $s_r$, adopting a planet-centered frame of reference introduces additional complexities. In particular, fluxes crossing the surface of a sphere centered on the planet do not necessarily correspond to fluxes directed toward the star. Therefore, to accurately calculate the fluxes moving toward the star, it is essential to perform surface integrations over spheres centered on the star rather than the planet.

\section{Increase of $S_r$ away from the planet\label{sec:appendixA}}
Figure \ref{fig:Sr_andSca_vs_r} clearly shows that the negative component of the surface-integrated quantity $S_r$, which represents the total power propagating toward the star through the surface, increases as one moves from the planet to the star (i.e., from right to left on the plot). Concurrently, the quantity $S \cdot \hat{c}_A$ exhibits a decreasing trend. This decrease is due to a gradual misalignment between the Poynting vector and the Alfvén characteristic with increasing distance from the planet. However, this trend does not fully explain the observed net increase in power directed toward the star, which continues to grow before eventually saturating. While the decrease in $S \cdot \hat{c}_A$ is noted, our focus here is not on its cause, as we have previously established that $S_r$ provides a more accurate measure of the inward power flow. Therefore, we turn our attention to investigating the physical mechanisms responsible for the amplification of the inward component of $S_r$ as it propagates from the planet toward the star.

We begin by decomposing the Poynting vector into its constituent components. To maintain clarity and avoid unnecessary complexity, constant factors such as $\pi$ and $\mu_0$ are omitted. consistent with the rest of the manuscript, lowercase `s' represent surface distributions and upper-case `S' represents surface integrated quantities. The derivation for the surface distribution proceeds as follows:

\begin{align}
\vec{s} &\equiv \vec{E} \times \vec{B} \\
        &\equiv -(\vec{v} \times \vec{B}) \times \vec{B} \\
        &\equiv \vec{v} B^2 - \vec{B} (\vec{v} \cdot \vec{B}) \\
\vec{s} \cdot \hat{r} 
        &\equiv [\vec{v} B^2] \cdot \hat{r} - [\vec{B} (\vec{v} \cdot \vec{B})] \cdot \hat{r} \\
s_r     &\equiv v_r B^2 - B_r (\vec{v} \cdot \vec{B}) \label{eq:Sr_decomposed}
\end{align}
\begin{figure}
    \centering
    \includegraphics[width=1.0\linewidth]{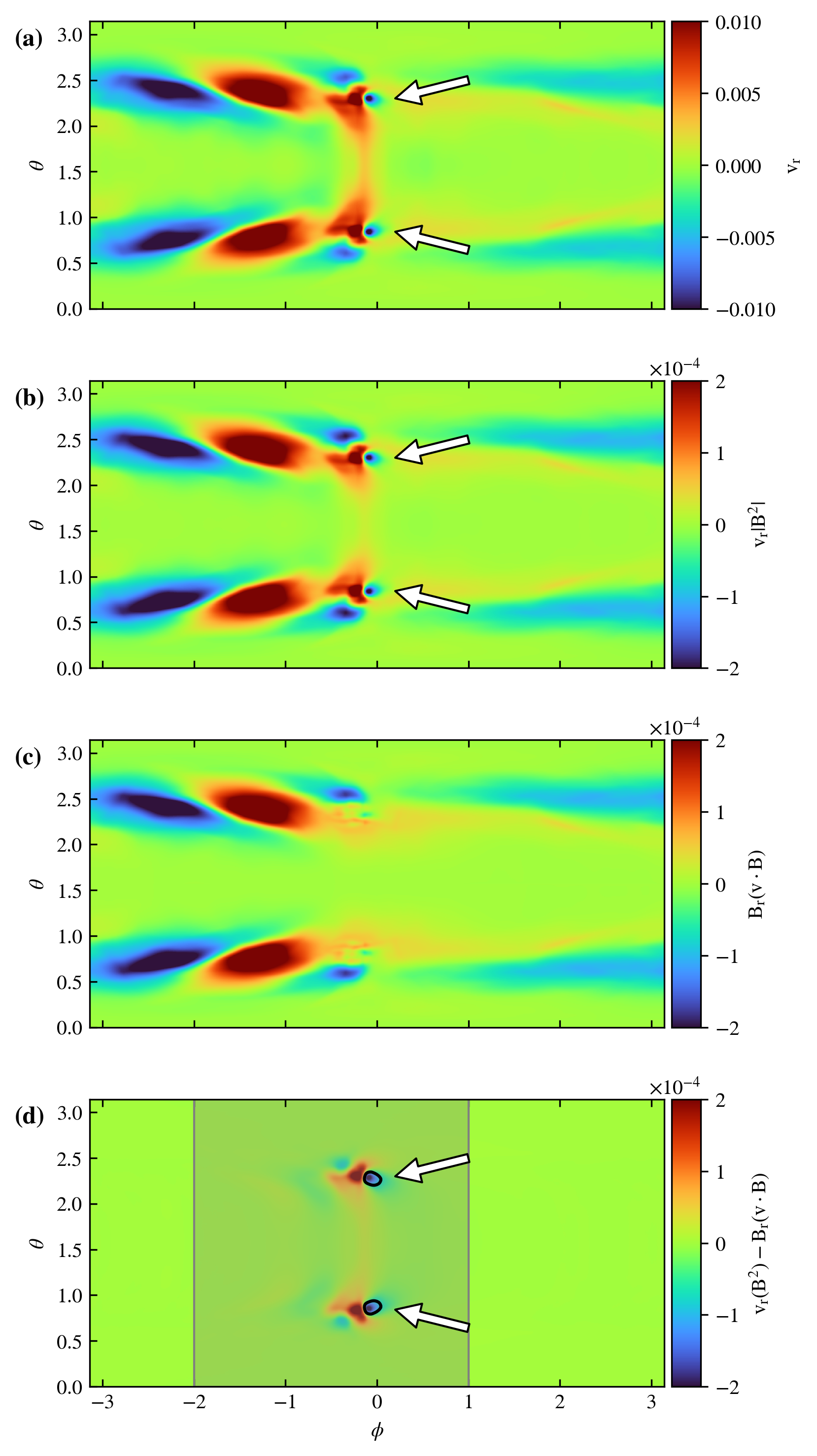}
    \caption{Radial velocity $v_r$ and components of the $s_r$ decomposition evaluated on a spherical surface centered on the star with radius $2.5,R_\star$; the planet in this setup orbits at $5,R_\star$. Panel (a) shows $v_r$; panels (b) and (c) show the two terms on the right-hand side of equation~\ref{eq:Sr_decomposed}; panel (d) shows their difference. The shaded region in panel (d) corresponds to the zoomed-in regions of Fig \ref{fig:bcomps_and_J}. All values are in code units.}
    \label{fig:vr_curl_comps}
\end{figure}

The final expression for $s_r$ (Eq \ref{eq:Sr_decomposed}), forms the foundation for analyzing the radial transport of electromagnetic energy. The right-hand side of this equation consists of two components. The first term, ($v_r B^2$), represents the advection of magnetic energy density ($B^2$) by the radial plasma velocity component $v_r$. The second term accounts for the reduction in energy transport due to plasma flow aligned with the magnetic field ($\vec{v}\cdot\vec{B}$); such field-aligned flow does not generate any motional electric field and consequently does not contribute to the Poynting vector. Together, these terms quantify the total electromagnetic energy flux passing through each radial integration surface illustrated in Figure \ref{fig:3d_schematic}. We note here that the entirety of the following analysis has been done in code units for simplicity.
Panel (a) of Figure \ref{fig:vr_curl_comps} shows the radial velocity $v_r$ evaluated on the same spherical surface as in Panel (a) and (b) of Figure \ref{fig:Sr_andSca_2D}, specifically at a radius of $2.5 R_\star$ in a system where the planet orbits at a distance of $5 R_\star$. The polar coordinate system used in this figure similarly represents the surface of a sphere centered on the star. The azimuthal angle $\phi = 0$ corresponds to the star–planet line, with $\phi$ ranging from $-\pi$ to $\pi$. The polar angle $\theta$ is defined such that $\theta = 0$ corresponds to the south stellar pole and increases to $\theta = \pi$ at the north stellar pole. Panels (b) and (c) show the two terms on the right-hand side of equation \ref{eq:Sr_decomposed} on the same surface. Panel (d) shows the difference between these two terms, which corresponds to $s_r$, and is therefore directly comparable to Panel (b) of Figure \ref{fig:Sr_andSca_2D}. Most of the large-scale features, both positive and negative, in the region $\phi < 0$, are present in both panels (b) and (c). However, since panel (d) is produced by subtracting panel (c) from panel (b), many of these overlapping features cancel out. The features of interest that remain prominent after subtraction are the negative lobes indicated by the white arrows in panel (b). These lobes initially appear to originate from regions of negative $v_r$ (however, as we discuss later, there are also contributions from the term $B^2$  making the term $v_rB^2$ the dominant component in this region), as shown in panel (a), where they are also marked by white arrows. Their persistence in the final result confirms that they are not removed by the subtraction process. Since these lobes are more intense in panel (b) in the negative values of $s_r$, we infer that they originate from the first term ($v_rB^2$) in equation \ref{eq:Sr_decomposed}. This demonstrates that the inward-directed electromagnetic energy flux (i.e., the negative portion of $s_r$) is directly associated with the advection of magnetic energy by radially inward plasma flow.

\begin{figure*}
    \centering
    \includegraphics[width=1.0\linewidth]{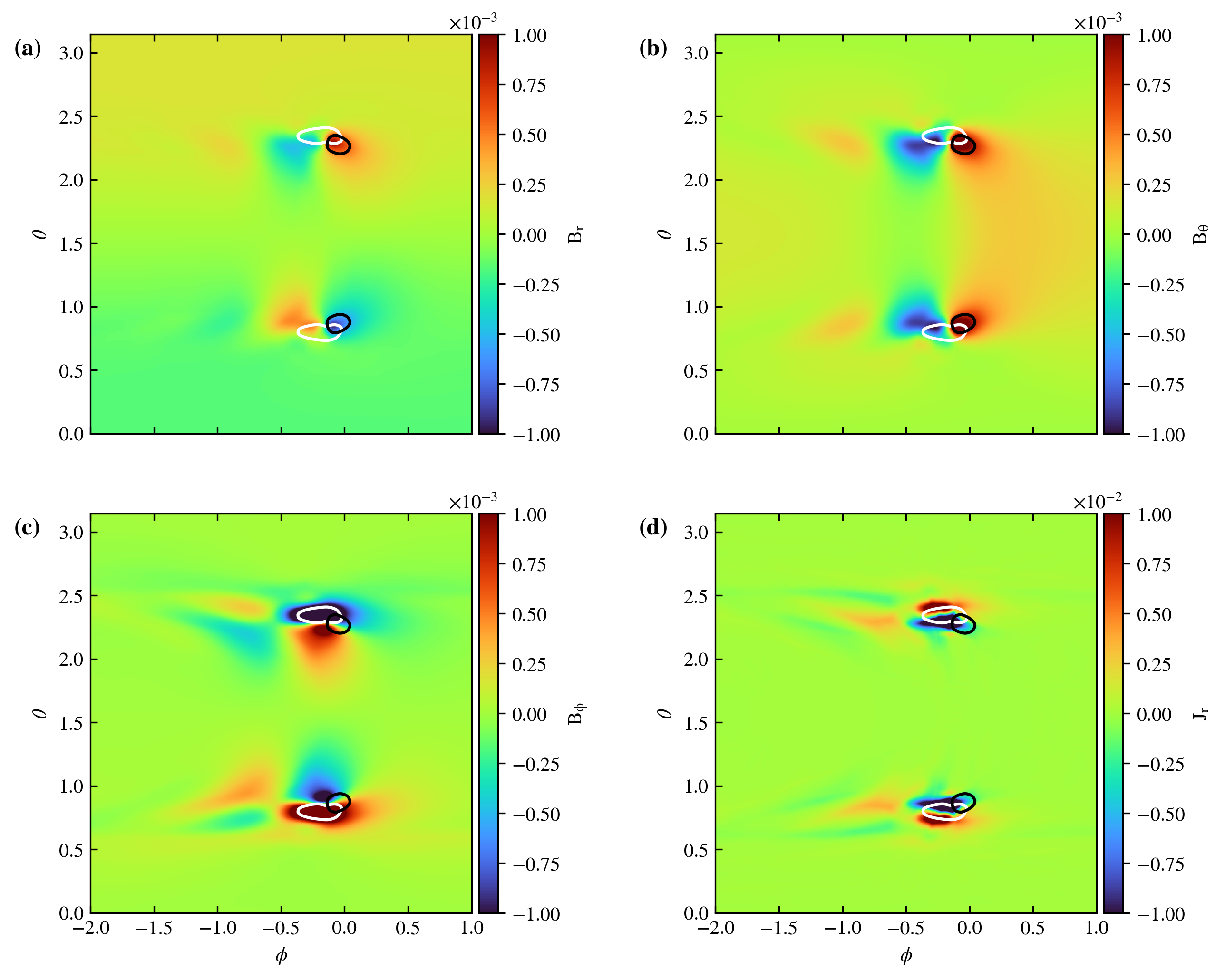}
    \caption{Spatial distribution of magnetic field components and current density on the same integration surface as in Figure \ref{fig:vr_curl_comps}. Panels (a), (b), and (c) show $B_r$, $B_\theta$, and $B_\phi$, respectively. The white contours represent cross sections of constant $\vec{s} \cdot \vec{\hat{c}}_A$, while the black contours denote the region where $s_r = -1 \times 10^{-4}$ in code units. Panel (d) shows the radial current density $J_r$, with strong current concentrations corresponding to the Alfvén wings.}
    \label{fig:bcomps_and_J}
\end{figure*}
Having established that the dominant contribution to the inward energy flux arises from the $v_r B^2$ term, we proceed to decompose this term into its individual components. Specifically, this can be expressed as $v_r |B^2| = v_r (B_r^2 + B_\theta^2 + B_\phi^2)$. Our analysis focuses on regions exhibiting strongly negative values of $S_r$, as these correspond to the areas most responsible for energy transfer toward the star. Panels (a), (b), and (c) of Figure \ref{fig:bcomps_and_J} present the spatial distribution of the magnetic field components on the same surface depicted in Figure \ref{fig:vr_curl_comps}. To facilitate detailed examination, the $\phi$ axis has been truncated to provide a zoomed-in perspective on the region of interest. Superimposed on these plots is a white contour delineating the cross section of constant $\vec{s} \cdot \vec{\hat{c}}_A$, which effectively maps the profile of the Alfvén wings. Additionally, black contours indicate levels of constant negative $s_r$, specifically at $s_r = -1 \times 10^{-4}$ in code units. It is immediately apparent that the regions of negative $s_r$ are localized predominantly at the leading edge of the Alfvén wings' cross section. Moreover, the spatial profile of the strongly inward-directed $s_r$ correlates closely with localized enhancements in both the $B_r$ and $B_\theta$ magnetic field components as seen in panels (a) and (b) of figure \ref{fig:bcomps_and_J}. It is important to emphasize that, although the magnetic field components themselves may assume positive or negative values, the quantity $B^2$ represents a squared magnitude and is therefore strictly positive.

Examining the $B_\phi$ component of the magnetic field, we observe that the saturated region of the color bar extends over a larger area, indicating a stronger overall magnitude compared to the $B_r$ and $B_\theta$ components shown in panels (a) and (b). Upon comparing magnitudes, the peak value of $B_\phi$ reaches approximately twice that of the maximum $B_\theta$ and about 3 times the maximum $B_r$ shown in the panels. However, it remains to be determined (and will be analyzed below) whether these strong magnetic field values contribute to the inflowing power. The white contour, representing the cross section of the Alfvén wings, closely follows the morphological structure of the $B_\phi$ enhancement. This strong spatial correlation further supports that the observed increase in the $B_\phi$ component is directly linked to the Alfvén wings. Indeed, when we plot the radial component of the current density ($J_r$) on the same surface in panel (d), it is evident that strong currents are concentrated within the Alfvén wings. This observation is consistent with previous studies, which have highlighted that Alfvén wings serve as carriers of significant current density \citep{Strugarek_2014}.  

\begin{figure}
    \centering
    \includegraphics[width=1.0\linewidth]{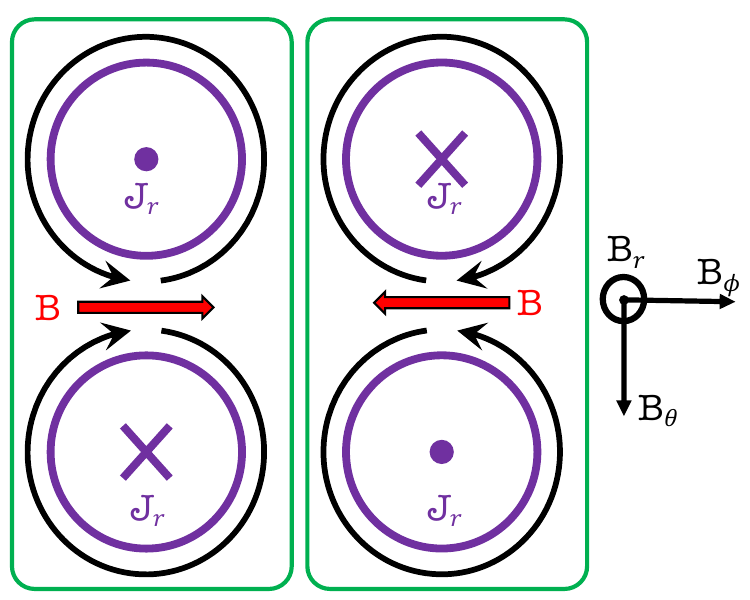}
    \caption{Schematic illustrating the magnetic field structure generated by bipolar radial currents, representative of those observed in panel (d) of Figure \ref{fig:bcomps_and_J}. The violet symbols ($\odot$ and $\otimes$) denote current flowing out of and into the plane, respectively. The black circular arrows depict the induced magnetic field loops. The superposition of these fields between the current poles is indicated by red arrows. The triad of spherical coordinate system unit vectors $(B_r, B_\theta, B_\phi)$ shown to the right of the figure.}
    \label{fig:JB_schematic}
\end{figure}

To investigate this behavior further, we present a schematic illustration aimed at clarifying the magnetic field structure generated by bipolar current patterns such as those observed in panel (d) of Figure \ref{fig:bcomps_and_J}. In Figure \ref{fig:JB_schematic}, violet symbols `$\odot$' and `$\otimes$' represent currents directed out of and into the plane, respectively. According to Ampère's law, these opposing radial currents produce loop-like magnetic fields, schematically depicted in the figure by black circular arrows. Notably, between the two current crossings, the superposition of these magnetic loops results in strong horizontal magnetic fields, as indicated by the red arrows. The coordinate system used in this analysis is shown in the right-hand side of the schematic and follows the standard spherical convention: $(r, \theta, \phi)$ extended to the context of the simulation. In this framework, the horizontal direction corresponds to the $\phi$ component. Therefore, this simple model provides a natural explanation for the pronounced $B_\phi$ enhancement observed in the cross section of the Alfvén wings.

\begin{figure}
    \centering
    \includegraphics[width=1.0\linewidth]{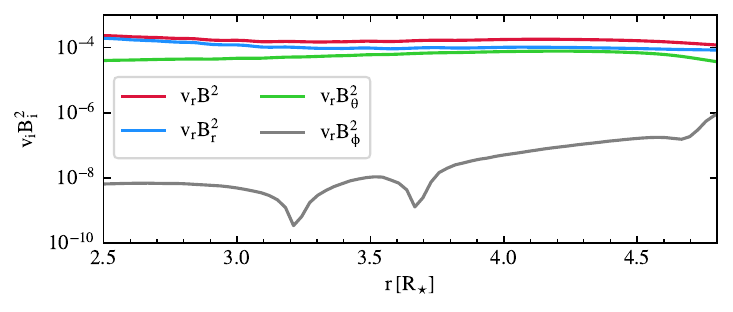}
    \caption{Surface integrated contributions of each component of ($v_r B^2$) to the radially inward Poynting flux as a function of radial distance from the star to the planet.}
    \label{fig:vBcomponents}
\end{figure}

Although there is a substantial enhancement of $B_\phi$ associated with the current systems within the Alfvén wings, this does not necessarily imply that it is the dominant contributor to the total $v_r B^2$ term. A closer inspection reveals that the regions of inward-directed $s_r$, indicated by black contours, predominantly lie at the periphery of the enhanced $B_\phi$ structure rather than overlapping with its core. To evaluate the relative contributions of the individual magnetic field components, we compute surface-integrated values of each term in $v_r(B_r^2 + B_\theta^2 + B_\phi^2)$, along with the total quantity. The integration is performed exclusively over regions where $s_r$ is negative, thereby isolating the contributions to the net inward-directed Poynting flux.

Figure \ref{fig:vBcomponents} presents the integrated values of each component as a function of radial distance from the star to the planet. The plot clearly demonstrates that the $v_r B_\phi^2$ contribution is several orders of magnitude smaller than those of $v_r B_r^2$ and $v_r B_\theta^2$. This confirms that the inward-directed Poynting flux is predominantly carried by the $B_r$ and $B_\theta$ components, while the enhanced $B_\phi$ contribution remains negligible in the overall energy transport toward the star.
\begin{figure}
    \centering
    \includegraphics[width=1.0\linewidth]{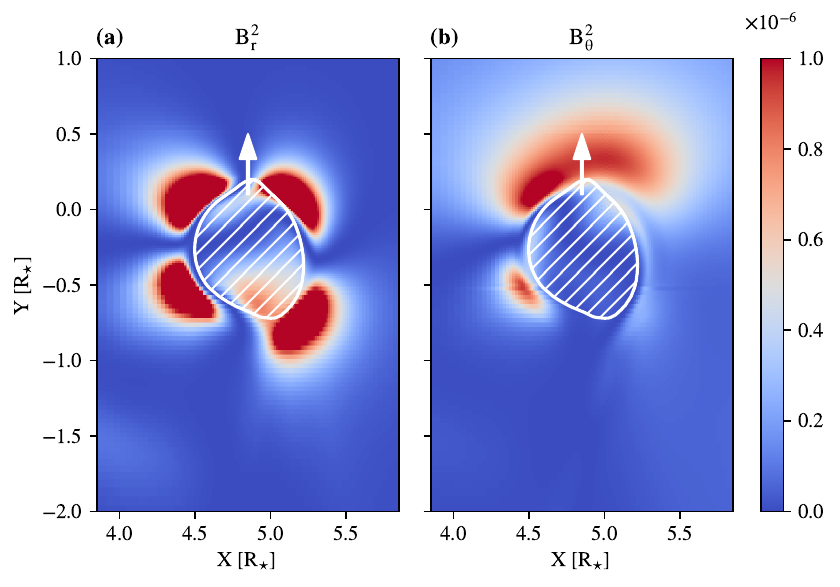}
    \caption{Panels (a) and (b) display 2D slices of the $B_r^2$ and $B_{\theta}^2$ components, respectively, in the X–Y plane just above the planetary obstacle. This plane lies parallel to the planet’s orbital plane and intersects the Alfvén wing (predominantly oriented along Z at this location) transversely. The slices correspond to $z = 0.7 R_{\star}$. The white hatched region indicates the cross section of the Alfvén wing, identified as the region comoving with the planet. The white arrows denote the direction of the orbital motion of the planet.}
    \label{fig:pileup}
\end{figure}
To dig deeper, we also plot the $B_r^2$ and $B_{\theta}^2$ components near the planet on an X-Y slice in figure \ref{fig:pileup}. This slice is taken just above the planetary body at a height of $z=0.7 R_{\star}$. The orbital distance of the planet is $5 R_\star$, so these plots correspond to the `r5b2' setup from Table \ref{tab:simulation_setups}. The white hatched regions indicate the cross sections of the Alfvén wings. These wings tether the planet to the star and therefore move together with the planet. Consequently, in the rest frame of the planet, the total plasma velocity inside the Alfvén wings is expected to be close to zero. The shaded region thus identifies where $|v| \sim 0$ in the orbital plane in this comoving frame as the cross section of the Alfvén wing \citep{Strugarek_2015}. The white arrows show the direction of the planet's orbital motion. Panels (a) and (b) of Figure \ref{fig:pileup} also reveal a clear enhancement of the magnetic fields at the leading edge of the Alfvén wing cross sections; both the $B_r^2$ and $B_{\theta}^2$ components display significant amplification there. Similar behavior is already observed in Figure \ref{fig:bcomps_and_J}, where the maximum inward Poynting flux occurs at the leading edge of the Alfvén wings. Additionally, panel (a) of Figure \ref{fig:vr_curl_comps} reveals an increase in $v_r$ in the same region as well, likely due to deflection of plasma by the Alfvén wings acting as extended obstacles. From these findings, we conclude that the enhancement of the Poynting flux arises from a combination of velocity deflection toward the star at the leading edge of the Alfvén wings and a simultaneous pileup of magnetic flux caused by the wings acting as an extended obstacle in the dynamic stellar wind. Together, these results suggest that the Alfvén wing magnetic field lines act as a large-scale obstacle to the stellar wind, generating Poynting flux directed toward the star along most of their length. Near the star, the relative velocity between the Alfvén wing field lines and the stellar wind plasma decreases significantly, reducing energy generation and causing saturation of the surface-integrated $S_r$ close to the star.

\section{Azimuthal confinement of SPMI power}
\begin{figure}
    \centering
    \includegraphics[width=1.0\linewidth]{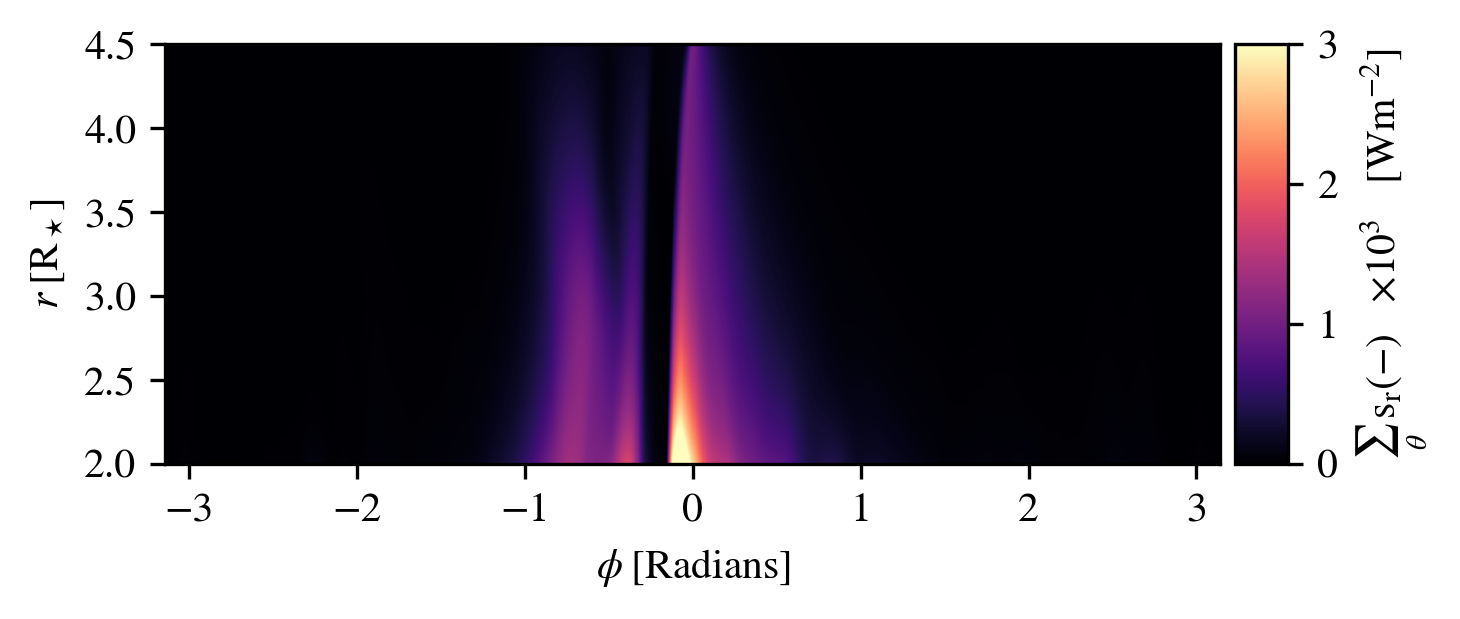}
    \caption{Longitudinal distribution of SPMI power on concentric spherical surfaces centered on the star. The y-axis indicates the radius of these spherical surfaces, while the x-axis shows the longitude. The color scale represents the latitude-integrated sum of the negative (inward-directed) Poynting flux in the northern hemisphere.}
    \label{fig:longitude_dep}
\end{figure}
Panel (b) of Figure \ref{fig:Sr_andSca_2D} demonstrates that the inward-directed power is localized in both latitude and longitude, supporting the concept of stellar hotspots formed by star-planet magnetic interactions (SPMIs). The prevailing understanding of stellar hotspots induced by SPMI holds that these hotspots move in phase with the exoplanet’s orbital period, which requires the energy deposition to be confined in longitude near the stellar surface. The exact longitude at which this power is concentrated strongly depends on the topology of the stellar magnetic field. \citet{Strugarek_2019} provides a detailed overview of the longitudinal dependence of energy deposition and its variation according to the stellar magnetic field lines to which the planet is connected. In the present study, due to the inherent symmetry where both the star and planet possess dipolar magnetic fields with their dipole axes aligned vertically to the orbital plane, the power deposition primarily occurs near the planet’s orbital longitude. To further illustrate that this longitudinal concentration persists along the entire Alfvén wing from the planet to the star, we present an additional plot showing the longitudinal distribution of SPMI power, obtained by integrating over latitude on spherical surfaces. This analysis is restricted to the Northern Hemisphere because the Southern Hemisphere exhibits symmetric behavior resulting from the computational domain’s symmetry. With the planet fixed at a longitude of zero radians, the plot reveals that the inward Poynting flux is strongly concentrated near this longitude, as expected from the non-tilted dipolar interaction topology. The power peak is slightly offset, trailing behind the subplanetary point, reflecting the Alfvén wing geometry where magnetic field lines curve toward negative longitudes due to the relative plasma flow before realigning with the planet’s orbital longitude closer to the star. We emphasize, however, that the longitudinal dependence of the Alfvén wings is strongly influenced by the magnetic field configuration of the SPMI system. Therefore, the result shown in Figure \ref{fig:longitude_dep} is specific to our simulation setup and may vary significantly for other magnetic field topologies.
\end{appendix}
\end{document}